\documentclass[prl,twocolumn,preprintnumbers,superscriptaddress,amsmath,amssymb]{revtex4-1}
\usepackage{graphicx}
\usepackage{subfigure}
\usepackage{mathrsfs}
\usepackage{amsfonts}
\usepackage{times}
\usepackage{amsmath}
\usepackage{leftidx}
\usepackage{color}
\usepackage[colorlinks,linkcolor=blue,citecolor=blue]{hyperref}
\usepackage{cleveref}

\begin{document}
\title{Topological Entanglement-Spectrum Crossing in Quench Dynamics}
\author{Zongping Gong}
\affiliation{Department of Physics, University of Tokyo, 7-3-1 Hongo, Bunkyo-ku, Tokyo 113-0033, Japan}
\author{Masahito Ueda}
\affiliation{Department of Physics, University of Tokyo, 7-3-1 Hongo, Bunkyo-ku, Tokyo 113-0033, Japan}
\affiliation{RIKEN Center for Emergent Matter Science (CEMS), Wako, Saitama 351-0198, Japan}
\date{\today}

\begin{abstract}
We unveil the stable $(d+1)$-dimensional topological structures underlying the quench dynamics for all the Altland-Zirnbauer classes in $d=1$ dimension, and propose to detect such dynamical topology from the time evolution of entanglement spectra. Focusing on systems in classes BDI and D, we find crossings in single-particle entanglement spectra for quantum quenches between different symmetry-protected topological phases. The entanglement-spectrum crossings are shown to be stable against symmetry-preserving disorder and faithfully reflect both $\mathbb{Z}$ (class BDI) and $\mathbb{Z}_2$ (class D) topological characterizations. As a byproduct, we unravel the topological origin of the global degeneracies emerging temporarily in the many-body entanglement spectrum in the quench dynamics of the transverse-field Ising model. These findings can experimentally be tested in ultracold atoms and trapped ions with the help of cutting-edge tomography for quantum many-body states. Our work paves the way towards a systematic understanding of the role of topology in quench dynamics.
\end{abstract}
\maketitle

\emph{Introduction.---}Topological quantum systems have attracted growing interest theoretically and experimentally \cite{Kane2010,Qi2011}, due partly to their fundamental importance in phase transitions beyond the conventional symmetry-breaking paradigm \cite{Wen2004} and applications to quantum computation \cite{Nayak2008,Sato2009,Sau2010,Alicea2011}. For gapped free-fermion systems at equilibrium,  a systematic classification has been established for the Altland-Zirnbauer (AZ) classes \cite{Altland1997,Ryu2008,Kitaev2009,Ryu2010,Teo2010} and with additional crystalline symmetries \cite{Ryu2013,Shiozaki2014,Fu2015,Shiozaki2016,Ryu2016}. Topological phases are characterized by topological invariants, some of which have been measured in ultracold atomic gases \cite{Monika2013b,Monika2015,Cooper2018a}. Entanglement measures \cite{Kitaev2006,Wen2006,Wen2009}, which are related to the full \emph{entanglement spectrum} (ES) \cite{Haldane2008,Pollmann2010,Fidkowski2010}, provide yet another powerful tool to detect topological order.

Recently, studies on topological systems have been extended to a nonequilibrium regime \cite{Eisert2015}. Floquet systems \cite{Moessner2017} have been demonstrated to exhibit intrinsically nonequilibrium topological phases with no static counterparts \cite{Kitagawa2010,Jiang2011,Lindner2013,Rudner2015,Lindner2016,Khemani2016,Potter2016,Vishwanath2016}. This Letter focuses on quantum quenches in topological systems \cite{Foster2013,Cooper2015,Vajna2015,Heyl2016,Balatsky2016,Budich2016,Refael2016,Cooper2016,Zhai2017,Weitenberg2017}. Starting from the ground state $|\Psi\rangle$ of an initial Hamiltonian $\hat H$, we suddenly change the Hamiltonian to $\hat H'$. The wave function subsequently undergoes a nontrivial unitary evolution $|\Psi(t)\rangle=e^{-i\hat H't}|\Psi\rangle$. Previous studies have unveiled topological dynamical phase transitions \cite{Vajna2015,Heyl2016,Balatsky2016}, a nonequilibrium Hall response which is not associated with the Chern number \cite{Budich2016,Refael2016,Cooper2016} and momentum-time Hopf links upon quenches during which the Chern number varies \cite{Zhai2017,Weitenberg2017}. Floquet quenches have also been investigated \cite{Rigol2015,Mitra2016,Mitra2017}.

However, it stays an open problem to \emph{systematically} identify and detect the topology of quench dynamics, i.e., the $(d+1)$-dimensional \emph{spatiotemporal} topology of the wave function. It is even unclear whether there is a \emph{stable} nontrivial $(d+1)$-dimensional dynamical topology that survives additional bands and disorder. Note that the Hopf link identified in Ref.~\cite{Zhai2017} is well-defined only for clean system with two bands \cite{Moore2008}. In this Letter, we demonstrate the existence of stable topological structures in quench dynamics and propose the time evolution of ES as their universal indicator. 
We use the K-theory to identify \emph{all} the AZ classes that accommodate stable nontrivial $(1+1)$-dimensional dynamical topology (see Table~\ref{table1}). We generalize the ES approach to quench dynamics and perform detailed model studies on topological systems in classes BDI and D, finding 
robust $\mathbb{Z}$ and $\mathbb{Z}_2$ topological features. Our study has strong relevance to state-of-the-art experiments of ultracold atoms \cite{Schmiedmayer2015,Daley2012,Greiner2015,Pichler2016,Bernien2017} and trapped ions \cite{Monroe2012,Monroe2016,Roos2017,Zhang2017b}, where many-body tomography has become possible \cite{Hauke2014,Weitenberg2016,Weitenberg2017,Weitenberg2018,Cramer2010,Blatt2017}. 
 
\emph{Parent Hamiltonian and its classification.---}For a formal classification, we note that the instantaneous many-body wave function $|\Psi(t)\rangle$ may be regarded as the ground state of 
\begin{equation}
\hat H(t)\equiv e^{-i\hat H't}\hat He^{i\hat H't}, 
\label{parHam}
\end{equation}
which we call the \emph{parent Hamiltonian}. Assuming that $\hat H$ and $\hat H'$ belong to the same $d$-dimensional AZ class, we obtain
\begin{equation}
\begin{split}
\hat{\mathcal{T}}\hat H(t)\hat{\mathcal{T}}^{-1}=\hat H(-t),&\;\;\;\;\;\;
\hat{\mathcal{C}}\hat H(t)\hat{\mathcal{C}}^{-1}=-\hat H(t),\\
\hat\Gamma\hat H(t)\hat\Gamma^{-1}&=-\hat H(-t),
\end{split}
\label{DynSym}
\end{equation} 
whenever $\hat H$ and $\hat H'$ respect  the time-reversal symmetry (TRS) $\hat{\mathcal{T}}$, the particle-hole symmetry (PHS) $\hat{\mathcal{C}}$ and/or the chiral symmetry $\hat\Gamma$. Regarding $t$ as an extra quasi-momentum, we find that $\hat H(t)$ respects the TRS in $d+1$ dimensions, but respects the PHS and/or the chiral symmetry \emph{only after reversing $t$ to $-t$}. Accordingly, 
the topological classification for $\hat H(t)$ subject to Eq.~(\ref{DynSym}) can differ qualitatively from that for the $(d+1)$-dimensional AZ classes \cite{Shiozaki2014}. For $d=1$, the results are shown in the second column in Table~\ref{table1} \cite{SMD}. We emphasize that the \emph{K-theory} classification, which has widely been applied to static topological insulators and superconductors \cite{Kitaev2009,Teo2010,Shiozaki2014,Shiozaki2016,Ryu2016}, places no constraints on the number of bands and the topology is expected to be robust against not-too-strong disorder \cite{Prodan2016,Thiang2016}. Here the disorder can stem not only from the absence of translation invariance in $\hat H$ and/or $\hat H'$, but also from the \emph{frequency domain} (Fourier transform of time) due to the band nonflatness in $\hat H'$ \footnote{The initial ground state stays unchanged if we flatten $\hat H$.}. 
These results can straightforwardly be generalized to quench dynamics in higher dimensions and/or with any additional two-fold symmetries \cite{Shiozaki2014}.

\begin{table}[tbp]
\caption{Topological classification of the parent Hamiltonians $\hat H(t)$ (\ref{parHam}) for quench dynamics. With symmetry constraints (\ref{DynSym}) alone, the classification is given by the maximal K-group, of which only a subset is dynamically realizable (third column).} 
\begin{center}
\begin{tabular}{ccc}
\hline\hline
Altland-Zirnbauer class & \;Maximal K-group\; & Dynamical realization \\
\hline
A & $\mathbb{Z}$ & 0 \\
AIII & $\mathbb{Z}\oplus\mathbb{Z}$ & $\mathbb{Z}$ \\
\hline
AI & 0 & 0 \\
BDI & $\mathbb{Z}$ & $\mathbb{Z}$ \\
D & $\mathbb{Z}_2$ & $\mathbb{Z}_2$ \\
DIII & $\mathbb{Z}_2\oplus\mathbb{Z}_2$ & $\mathbb{Z}_2$ \\
AII & $\mathbb{Z}_2$ & 0 \\
CII & $\mathbb{Z}$ & $\mathbb{Z}$ \\
C & 0 & 0 \\
CI & 0 & 0 \\
\hline\hline
\end{tabular}
\end{center}
\label{table1}
\end{table}

At this stage, it is unclear whether a nontrivial element in these 
maximal K-groups 
can be realized by parent Hamiltonians (\ref{parHam}), which has a specific $t$ dependence. After a one-by-one examination \cite{SMD}, we identify all the dynamically realizable elements 
in the third column in Table~\ref{table1}. It turns out that, for all the nontrivial AZ classes, the \emph{two-dimensional} topological index, i.e., \emph{the strong topological number} \cite{Kitaev2009} of $\hat H(t)$ in Eq.~(\ref{parHam}), is simply the difference between the one-dimensional topological indices of $H$ and $H'$. This is why the results coincide with the one-dimensional column in the well-known periodic table \cite{Altland1997,Ryu2008,Kitaev2009,Ryu2010}. For example, the strong topological number $\mathbb{Z}$ of $\hat H(t)$ in class BDI is given by
\begin{equation}
\Delta w=w'-w,
\end{equation}
where $w$ ($w'$) is the winding numbers of $\hat H$ ($\hat H'$). The $\mathbb{Z}_2$ index $v$ of $\hat H(t)$ in class D, as first identified in adiabatic PHS-protected pumps \cite{Teo2010}, is given by
\begin{equation} 
v=|\mathcal{N}'-\mathcal{N}|,
\end{equation}
where $\mathcal{N}$ ($\mathcal{N}'$) is the $\mathbb{Z}_2$ index of $\hat H$ ($\hat H'$). We will illustrate these two classes with concrete models. 

Two remarks are in order. First, the topological numbers in the maximal K-groups which are absent in quench dynamics can take nonzero values in adiabatic topological pumps \cite{Thouless1983,Fu2006,Monika2016b,Takahashi2016}. Second, the \emph{weak topological numbers} \cite{Kitaev2009} of lower-dimensional nature are not shown in Table~\ref{table1}. In fact, the conserving Chern number in quench dynamics in two dimensions found in Refs.~\cite{Rigol2015,Cooper2015} gives such an example. Here, we find another example --- the $\mathbb{Z}_2$ index of one-dimensional systems in class D. In other classes, however, the one-dimensional topological index may change 
in quench dynamics \cite{Cooper2018b}.   

\emph{Entanglement-spectrum dynamics after quench.---}With the topology of quench dynamics formally identified, it is natural to ask how to detect it in a way that is universal, numerically tractable and experimentally accessible. For static free-fermion systems $\hat H=\sum_{j,l,\alpha,\beta}H_{j\alpha,l\beta}\hat c^\dag_{j\alpha}\hat c_{l\beta}$, where $\hat c^\dag_{j\alpha}$ creates a particle with internal degrees of freedom $\alpha$ at site $j$, an ideal candidate is the \emph{single-particle ES}, which gives the \emph{exact} open-boundary spectrum of the flattened Hamiltonian \cite{Fidkowski2010}. As for quench dynamics, the \emph{time evolution of ES} thus faithfully simulates the edge spectrum flow under open-boundary conditions in real space. Given the \emph{bulk-edge correspondence} \cite{Prodan2016}, we expect that the dynamical topology can directly be readout from the ES dynamics. Note that the converse use of this idea can be practically useful for recovering the Hamiltonian topology from quench dynamics, provided that the many-body tomography for $|\Psi(t)\rangle$ \cite{Cramer2010,Hauke2014} or the direct measurement of the ES \cite{Pichler2016,Dalmonte2018} is achievable.

We sketch out the definition of the single-particle ES of a Gaussian state $|\Psi\rangle$. Denoting $S$ ($\bar S$) as the region of interest (the complementary of $S$), the reduced density operator $\hat\rho_S\equiv{\rm Tr}_{\bar S}[|\Psi\rangle\langle\Psi|]$ can be rewritten as $\hat\rho_S=Z^{-1}_{\rm E}e^{-\hat H_{\rm E}}$, with $\hat H_{\rm E}=\sum_n\epsilon_n\hat f^\dag_n\hat f_n$ being the quadratic entanglement Hamiltonian \cite{Peschel2003}, where $\hat f_n$ is linear in $\hat c_{j\alpha}$. The single-particle ES is given by \cite{Hughes2011} \begin{equation}
\xi_n\equiv\frac{1}{e^{\epsilon_n}+1}, 
\label{spES}
\end{equation}
so that an entanglement zero mode $\epsilon_n=0$ corresponds to $\xi_n=\frac{1}{2}$. To investigate the ES dynamics, we calculate $\xi_n$ for $|\Psi(t)\rangle$ at each time in concrete models in classes BDI and D and visualize the $\mathbb{Z}$ and $\mathbb{Z}_2$ indices.

\begin{figure*}
\begin{center}
        \includegraphics[width=17cm, clip]{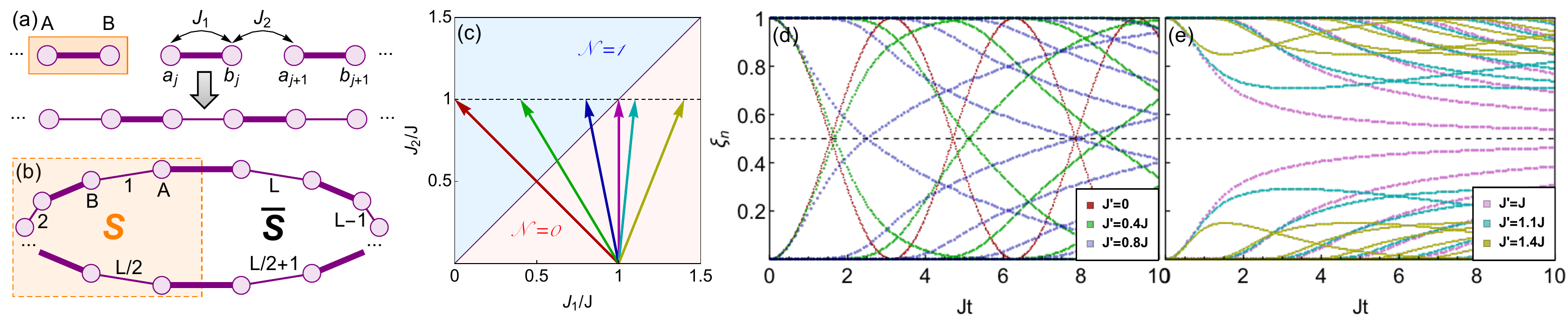}
      \end{center}
   \caption{(color online) (a) Quench in the SSH model (\ref{HSSH}) 
   from a dimerized state. The orange rectangle marks a unit cell. (b)  Half-chain entanglement cut (shaded region $S$) of a periodic chain. (c) Quench protocols. The leftmost three arrows show quenches across the topological phase boundary. (d) Dynamics of the single-particle ES (\ref{spES}) after quenches across the phase boundary, showing crossings at $\xi_n=\frac{1}{2}$. The total number of $\xi_n$'s is $L$ and most of them are very close to $0$ or $1$. (e) 
   Single-particle ES dynamics after quenches within the same phase and to the critical point, showing no crossings at $\xi_n=\frac{1}{2}$. The system size is $L=100$.}
   \label{fig1}
\end{figure*}

\emph{Two-band BDI systems in one dimension.---}We start with 
two-band systems in class BDI. Without loss of generality, we denote the Bloch Hamiltonian as $h(k)=\boldsymbol{d}(k)\cdot\boldsymbol{\sigma}$, where $\boldsymbol{\sigma}\equiv\sum_{\mu=x,y,z}\sigma^\mu\boldsymbol{e}_\mu$ is the Pauli-matrix vector with $\boldsymbol{e}_\mu$ being the unit vector in the $\mu$ direction. The Hamiltonian $\hat H$ can be related to $h(k)$ by $\hat H=\sum_k\hat{\boldsymbol{c}}^\dag_kh(k)\hat{\boldsymbol{c}}_k$, where $\hat{\boldsymbol{c}}_k\equiv(\hat a_k,\hat b_k)^{\rm T}$, $\hat a_k\equiv\frac{1}{\sqrt{L}}\sum_je^{-ikj}\hat a_j$ ($\hat b_k\equiv\frac{1}{\sqrt{L}}\sum_je^{-ikj}\hat b_j$), $L$ is the number of unit cells and $\hat a_j$ ($\hat b_j$) annihilates a fermion in the A (B) sublattice in the $j$th unit cell (see Fig.~\ref{fig1}(a)).

Now we impose TRS $\hat{\mathcal T}$ and PHS $\hat{\mathcal C}$, which satisfy $\hat{\mathcal T}^2=\hat{\mathcal C}^2=1$, $\hat{\mathcal T}\hat{\boldsymbol{c}}_k\hat{\mathcal T}^{-1}=\hat{\boldsymbol{c}}_{-k}$ and $\hat{\mathcal C}\hat{\boldsymbol{c}}_k\hat{\mathcal C}^{-1}=\sigma^z\hat{\boldsymbol{c}}_{-k}$. In terms of the $\boldsymbol{d}$-vector, the symmetry constraints $[\hat H,\hat{\mathcal{T}}]=\{\hat H,\hat{\mathcal{C}}\}=0$ imply $d_x(k)=d_x(-k)$, $d_y(k)=-d_y(-k)$ and $d_z(k)=0$. Note that $[\sigma^x,h(\Gamma)]=0$ at high-symmetry points $\Gamma=0,\pi$, where the Bloch state is an eigenstate of $\sigma^x$ with eigenvalue $\nu_\Gamma=\pm1$. The winding number is determined by $w\equiv\int^\pi_{-\pi}\frac{dk}{2\pi}\frac{q'(k)}{q(k)}$ with $q(k)\equiv d_x(k)-id_y(k)$, and the PHS-protected $\mathbb{Z}_2$ index reads $\mathcal{N}\equiv\frac{1}{2}|\nu_0-\nu_\pi|=w\;{\rm mod}\;2$. 

A prototypical example in class BDI is the Su-Schrieffer-Heeger (SSH) model \cite{Su1979}:
\begin{equation}
\hat H=-\sum_j(J_1\hat b^\dag_j\hat a_j+J_2\hat a^\dag_{j+1}\hat b_j+{\rm H.c.}),
\label{HSSH}
\end{equation}
where $J_1$ and $J_2$ are the intra- and inter-unit-cell hopping amplitudes, respectively. The Fourier transform of Eq.~(\ref{HSSH}) gives $\boldsymbol{d}(k)=-(J_1+J_2\cos k,J_2\sin k,0)$, implying $(\nu_0,\nu_\pi)=(\frac{J_1+J_2}{|J_1+J_2|},\frac{J_1-J_2}{|J_1-J_2|})$. In real 
systems such as polyacetylene \cite{Shirakawa1977} and ultracold atoms 
\cite{Monika2013b,Monika2016b,Takahashi2016}, we generally have $J_1,J_2>0$, and a topological phase transition from $\mathcal{N}=0$ to $\mathcal{N}=1$ occurs upon crossing the boundary $J_1=J_2$ (see Fig.~\ref{fig1}(c)).

If we quench the parameters in the SSH model (\ref{HSSH}) as 
$(J_1,J_2)=(J,0)\to(J',J)$, 
$|\Psi(t)\rangle$ will remain in the same trivial phase as the dimerized state with $\mathcal{N}=0$. Hence, topological entanglement edge modes in $|\Psi(t)\rangle$ are absent 
in general. This is confirmed numerically, i.e., the half-chain (see Fig.~\ref{fig1} (b)) ES $\xi_n\neq\frac{1}{2}$ for almost all the time in Figs.~\ref{fig1}(d) and (e). However, in the flat-band case $J'=0$, we find periodic oscillations of $\xi_n$'s, which cross each other at $t_m=(m-\frac{1}{2})\frac{\pi}{J}$ with $m\in\mathbb{Z}^+$, where the system instantaneously becomes class BDI with winding number $2$. 
Remarkably, the crossings stay robust as $J'$ increases as long as $J'<J$ with $t_1$ gradually diverging. This should be understood as the robustness of the nontrivial $(1+1)$-dimensional topology characterized by $\Delta w=1$, although the temporal periodicity disappears. 
When $J$ exceeds $J'$, no crossings occur. 
This sharp transition in the ES dynamics 
distinguishes the 
quenches across different topological phases from those within the same phase.

The ES crossings can alternatively be interpreted as a result of the nontrivial PHS-protected index $v=1$, which equals the Skyrmion charge (Chern number) of the $\boldsymbol{d}$-vector textures in one half of the momentum-time space \cite{SMD}. Indeed, the ES crossings resemble the Dirac-cone dispersion of edge (entanglement) modes in two-dimensional topological insulators \cite{Kane2005,Vishwanath2010}. 
The Chern number can be nonzero since $d_z$ is dynamically generated even if it vanishes in both $h(k)$ and $h'(k)$. Such a dynamical Chern number is recently identified for general two-band systems \cite{Yang2018}, and should be experimentally measurable with the help of Bloch-state tomography for ultracold atoms in optical lattices \cite{Hauke2014,Weitenberg2016,Weitenberg2017,Weitenberg2018}.


\emph{Influence of the band number, disorder, and symmetry breaking.---}In the presence of additional bands and/or disorder, the picture of momentum-time Skyrmions mentioned above breaks down and only a $\mathbb{Z}_2$ index instead of a PHS-protected Chern number is well-defined. Nevertheless, we will show that the ES dynamics stays a good indicator for the stable dynamical topology and clearly distinguishes the $\mathbb{Z}$ (class BDI) characterization from the $\mathbb{Z}_2$ (class D) one.

\begin{figure}
\begin{center}
        \includegraphics[width=8.5cm, clip]{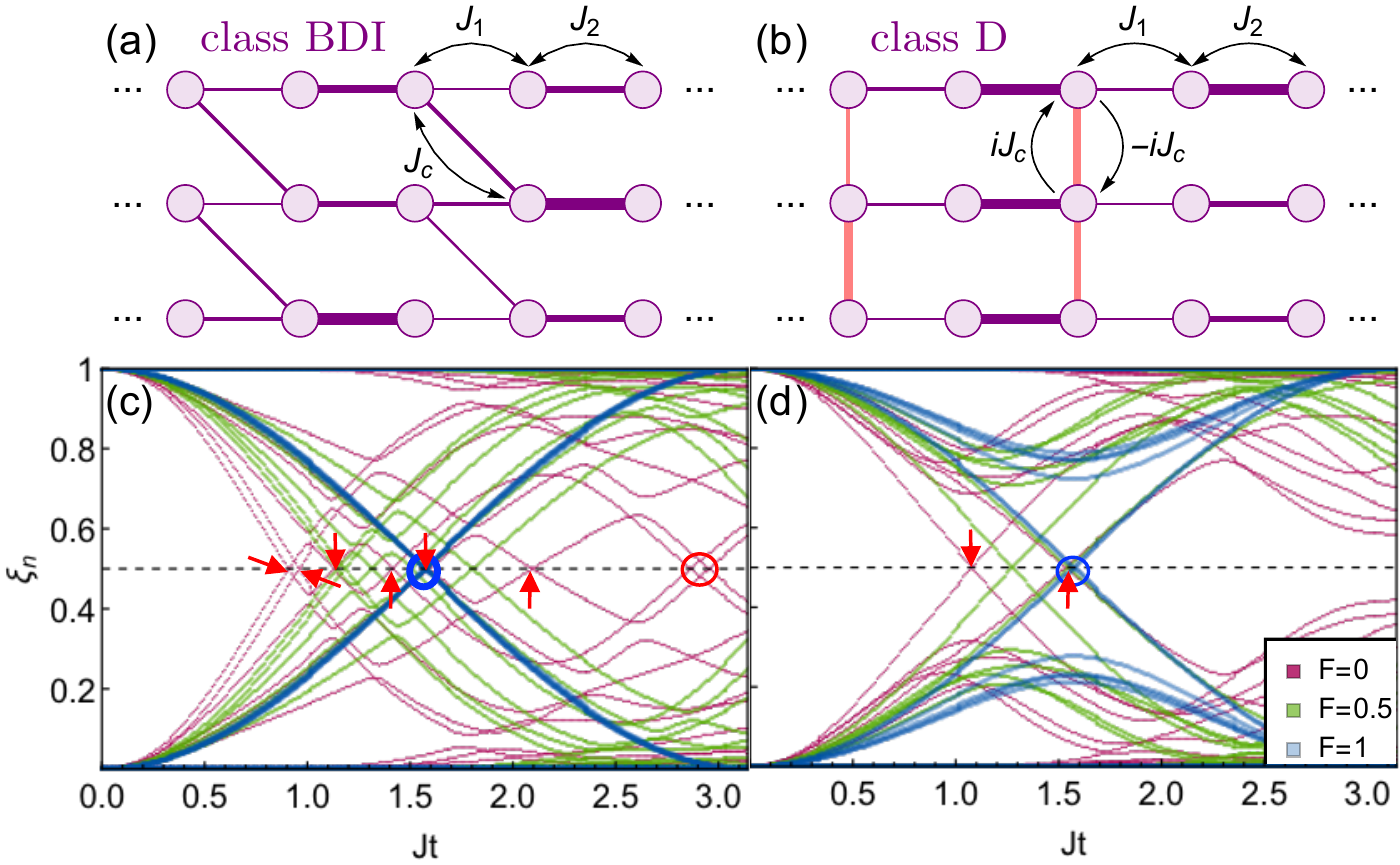}
      \end{center}
   \caption{(color online) Three coupled SSH chains in (a) class BDI and (b) class D. Hopping amplitudes $J_\alpha$ ($\alpha=1,2,c$) are randomly sampled from a uniform distribution over $[0.6\bar J_\alpha,1.4\bar J_\alpha]$. (c) ES dynamics after quench $(\bar J_1,\bar J_2,\bar J_c)=(0,1.5J,0)\to(1.5J,0.5J,0.5J)$ in (a) with $L=40$ and the periodic-boundary condition. The result ($F=0$) is compared with those after partial ($F=0.5$) and complete ($F=1$) band flattening $\hat H'$. A partially flattened Hamiltonian $\hat H'_F$ ($F\in(0,1)$) is related to the original one $\hat H'_0=\hat H'$ and the completely flattened one $\hat H'_1$ via $\hat H'_F=F\hat H'_1+(1-F)\hat H'_0$. The ES crossings in the blue circle split into those marked by red arrows when $F$ changes from $1$ to $0$. The remaining two crossings in the red circle stem from the second period. (d) Same as (c) but for the system in (b) with a different quench protocol $(\bar J_1,\bar J_2,\bar J_c)=(0,1.5J,0)\to(1.5J,0.5J,J)$.}
   \label{fig2}
\end{figure}

According to Table~\ref{table1}, the quench dynamics in class BDI systems are characterized by $\mathbb{Z}$. Since the addition operation on a K-group is the direct sum up to continuous deformation, we expect the number of ES crossings to be multiplied by $M$ if we quench $M$ copies of the system coupled to each other without breaking the symmetries (see Fig.~\ref{fig2}(a)), provided that the disorder in the frequency domain due to band nonflatness is not so strong. We numerically confirm this for $M=1\sim4$ SSH chains with hopping disorder \cite{SMD}. An example for $M=3$ is shown in Fig.~\ref{fig2}(c), where we see $2M=6$-fold degenerate ES crossings in the flat-band limit, with the factor of $2$ coming from the periodic-boundary condition. 
Note that the crossings for nonflat bands are more like middle-gap edge states, a feature well-known in topological crystalline systems \cite{Hughes2011}. Indeed, $\hat{\mathcal{C}}$ behaves like a crystalline symmetry. 

If we break TRS alone, the symmetry class changes from BDI to D and the K-theory classification gives $\mathbb{Z}_2$ (see Table~\ref{table1}), over which $1_{\mathbb{Z}_2}+1_{\mathbb{Z}_2}=0_{\mathbb{Z}_2}$. As a result, we expect the presence (absence \footnote{We do not rule out the possibility of finding accidental ES crossings without topological origin. Note that edge states may also exist in topologically trivial systems.}) of ES crossings if we quench an odd (even) copies of  SSH chains with coupling amplitudes respecting PHS but breaking TRS (see Fig.~\ref{fig2}(b)). In Fig.~\ref{fig2}(d), we present the results for $M=3$ chains. We find that only a single pair of crossings survive in a period in the flat-band limit, and the crossings persist when introducing band nonflatness. We have observed a similar behavior in class DIII \cite{SMD}, which is also characterized by $\mathbb{Z}_2$ (see Table~\ref{table1}).

\begin{figure}
\begin{center}
        \includegraphics[width=8.5cm, clip]{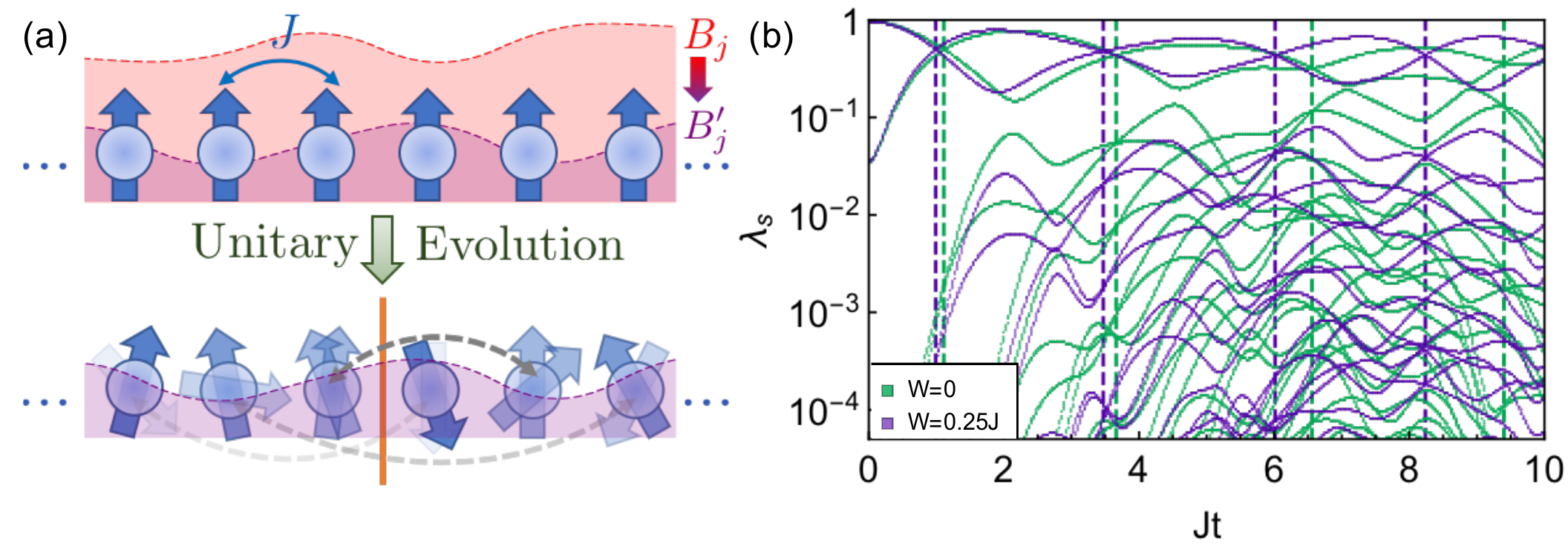}
      \end{center}
   \caption{(color online) (a) With an 
   inhomogeneous magnetic field quenched, a nearly disentangled paramagnetic Ising chain (\ref{TFI}) becomes entangled under unitary evolution. The orange line denotes the half-chain 
   cut. (b) Dynamics of the many-body ES $\lambda_{\boldsymbol{s}}$ (\ref{mbl}). Quench protocol: $(J,\bar B_j)=(1,1.5)\to(1,0.5)$, with ($W=0.25J$) or without disorder ($W=0$). The length of the open Ising chain is $L=10$. The dashed lines indicate where global two-fold degeneracies emerge.}
   \label{fig3}
\end{figure}

\emph{Discussions.---}The ES dynamics 
has been discussed in the transverse-field Ising model \cite{Chiara2014}, which can be mapped to the Kitaev chain \cite{Kitaev2001}. Therein, 
\emph{global} two-fold degeneracies emerge in the many-body ES $\lambda_{\boldsymbol{s}}$'s at certain times upon the field quench across the critical value. Since the many-body ES $\{\lambda_{\boldsymbol{s}}\}$ as eigenvalues of $\hat\rho_S$ are related to $\xi_n$'s via \cite{Fidkowski2010}
\begin{equation}
\lambda_{\boldsymbol{s}=\{s_n\}}=\prod_n\left[\frac{1}{2}+s_n\left(\xi_n-\frac{1}{2}\right)\right],\;\;\;\;s_n=\pm1,
\label{mbl}
\end{equation}
we can attribute these global degeneracies to single-particle ES crossings at $\frac{1}{2}$. Since the Kitaev chain belongs to class D, according to Table~\ref{table1}, we expect the global many-body ES degeneracies 
to be robust against disorder. This is confirmed in 
an Ising chain subject to an 
inhomogeneous magnetic field:
\begin{equation}
\hat H=\sum_j(J\hat\sigma^x_j\hat\sigma^x_{j+1}+B_j\hat\sigma^z_j),
\label{TFI}
\end{equation}
where $B_j$ obeys a uniform distribution over $[\bar B_j-W,\bar B_j+W]$. As shown in Fig.~\ref{fig3}(b), the global many-body ES degeneracies persist in spite of disorder, although they appear at different times. 
We have further checked the robustness against random coupling \cite{SMD}. Such a topological dynamical phenomena can be explored in trapped-ion systems with the help of matrix-product-state tomography \cite{Cramer2010,Blatt2017}.

It was conjectured \cite{Chiara2014} that the emergence of many-body-ES degeneracies is related to a dynamical quantum phase transition \cite{Heyl2013} associated with singularities of the dynamical free-energy density $f(t)\equiv-\lim_{L\to\infty}\frac{1}{L}\ln|\langle\Psi|e^{-i\hat H't}|\Psi\rangle|^2$. As for the SSH model, everytime $f(t)$ becomes nonanalytic, we arrive at the center of a momentum-time Skyrmion. However, a precise numerical analysis indicates that these times do not exactly coincide with those of ES crossings 
\cite{SMD}. Furthermore, a dynamical phase transition may occur without ES crossings in the Rice-Mele model \cite{Rice1982}. Therefore, dynamical phase transitions and ES crossings are not equivalent, although there could be a sufficient condition for both \cite{Balatsky2016}. Similar conclusions are drawn in Ref.~\cite{Zhai2017} for quench dynamics in two dimensions.

The ES dynamics has also been studied in the context of topological Floquet systems \cite{Potter2016,Yao2017b}. 
A prototypical example of a 
modulated Ising chain is studied in Ref.~\cite{Potter2016}, which is reminscient of a quench $\hat H=\sum_jB_j\hat\sigma^z_j\to\hat H'=J\sum_j\hat\sigma^x_j\hat\sigma^x_{j+1}$ in a single period. However, in Ref.~\cite{Potter2016}, the ES dynamics is for \emph{Floquet eigenstates} and the robustness of crossing is discussed through perturbations with the same Floquet period; here we focus on physical states undergoing unitary evolution generated by time-independent Hamiltonians and the temporal periodicity is generally absent.

\emph{Summary and outlook.---}We have identified the stable topological structures for all the one-dimensional quench dynamics within the same AZ class. We have proposed using the ES dynamics to detect the dynamical topology and performed detailed model studies for classes BDI and D. We have numerically demonstrated the robust $\mathbb{Z}$ and $\mathbb{Z}_2$ features. These phenomena can be explored in state-of-the-art ultracold-atom and trapped-ion experiments \cite{SMD}. 

In higher dimensions \cite{Chang2018} and/or with additional symmetries, there remains an open problem as to whether a nontrivial $(d+1)$-dimensional topological structure emerges in quench dynamics, and, if yes, how the single-particle ES dynamics looks like. The influence of interaction is another important issue, which might be tackled from the dynamics of the many-body ES. In one dimension, this can be readout from the matrix-product-state representation \cite{Vidal2007}. Examples are provided in Supplemental Material \cite{SMD}.


We acknowledge K. Shiozaki, S. Furukawa, M. Sato, S. Higashikawa, Y. Ashida and M. Nakagawa for valuable discussions. This work was supported by KAKENHI Grant No. JP18H01145 from the Japan Society for the Promotion of Science, a Grant-in-Aid for Scientic Research on Innovative Areas ``Topological Materials Science" (KAKENHI Grant No. JP15H05855), and the Photon Frontier Network Program from MEXT of Japan, and the Mitsubishi Foundation. Z. G. was supported by MEXT.

\bibliography{GZP_references}

\clearpage
\begin{center}
\textbf{\large Supplemental Materials}
\end{center}
\setcounter{equation}{0}
\setcounter{figure}{0}
\setcounter{table}{0}
\makeatletter
\renewcommand{\theequation}{S\arabic{equation}}
\renewcommand{\thefigure}{S\arabic{figure}}
\renewcommand{\bibnumfmt}[1]{[S#1]}

Here we provide the details of the K-theory classification of quench dynamics in one dimension, the details on the calculation of the entanglement-spectrum dynamics, the detailed calculations for two-band models, a comment on the relationship of the entanglement spectrum to the dynamical phase transition, and a sketch of experimental situations.

\section{K-theory analysis of quench dynamics}

In this section, we employ the K-theory \cite{Kitaev2009,Teo2010,Shiozaki2014} to analyze possible nontrivial topological classifications for the parent Hamiltonian 
\begin{equation}
\hat H(t)=e^{-i\hat H't}\hat He^{i\hat H't},
\label{Ht}
\end{equation}
where the pre- and postquench Hamiltonians $\hat H$ and $\hat H'$ are assumed to belong to the same Altland-Zirnbauer (AZ) class.

\subsection{Parent Bloch Hamiltonian}
We first show that, in the presence of translation invariance, the parent Bloch Hamiltonian $h(k,t)$ defined from $\hat H(t)=\sum_k\hat{\boldsymbol{c}}^\dag_kh(k,t)\hat{\boldsymbol{c}}_k$ is given by
\begin{equation}
h(k,t)=e^{-ih'(k)t}h(k)e^{ih'(k)t},
\label{parBlochHam}
\end{equation} 
where $h(k)$ ($h'(k)$) is the Bloch Hamiltonian of $\hat H$ ($\hat H'$), i.e., $\hat H=\sum_k\hat{\boldsymbol{c}}^\dag_kh(k)\hat{\boldsymbol{c}}_k$ ($\hat H'=\sum_k\hat{\boldsymbol{c}}^\dag_kh'(k)\hat{\boldsymbol{c}}_k$). Here $\hat{\boldsymbol{c}}_k$ is a vector consisting of operators $\hat c_{k\alpha}$'s with different $\alpha$, where $k$ is the wave number and $\alpha$ labels the internal degrees of freedom, such as sublattices and spins.

The equation of motion for $\hat H(t)$ reads
\begin{equation}
i\partial_t\hat H(t)=[\hat H',\hat H(t)],\;\;\;\;\hat H(0)=\hat H.
\label{HeisenH}
\end{equation}
With translation invariance, Eq.~(\ref{HeisenH}) can be decomposed into independent blocks having different quasimomenta:
\begin{equation}
\begin{split}
i\partial_t\hat{\boldsymbol{c}}^\dag_kh(k,t)\hat{\boldsymbol{c}}_k&=[\hat{\boldsymbol{c}}^\dag_kh'(k)\hat{\boldsymbol{c}}_k,\hat{\boldsymbol{c}}^\dag_kh(k,t)\hat{\boldsymbol{c}}_k]\\
&=\hat{\boldsymbol{c}}^\dag_k[h'(k),h(k,t)]\hat{\boldsymbol{c}}_k,
\end{split}
\label{Block}
\end{equation}
with the initial condition $h(k,0)=h(k)$. Here we have used the fermion-operator identity:
\begin{equation} 
[\hat c^\dag_{k\alpha}\hat c_{k\beta},\hat c^\dag_{k\gamma}\hat c_{k\delta}]=\delta_{\beta\gamma}\hat c^\dag_{k\alpha}\hat c_{k\delta}-\delta_{\alpha\delta}\hat c^\dag_{k\gamma}\hat c_{k\beta}.
\end{equation}
Note that Eq.~(\ref{Block}) implies Eq.~(\ref{parBlochHam}). We mention at this stage 
the spatial dimension or the total number of bands need not be specified.

While $h(k,t)$ in Eq.~(\ref{parBlochHam}) is not periodic in $t$ in general, we can always flatten $h(k)$ and $h'(k)$ into $h_1(k)$ and $h'_1(k)$, which satisfy $h^2_1(k)=h'^2_1(k)=J^2$ and share the same symmetry as $h(k)$ and $h'(k)$ \cite{Kitaev2009}. The corresponding flattened parent Bloch Hamiltonian takes the form
\begin{equation}
\tilde h(k,t)=e^{-i h'_1(k)t}h_1(k)e^{i h'_1(k)t},
\label{flatPBH}
\end{equation}
which satisfies $\tilde h(k,t+\frac{\pi}{J})=\tilde h(k,t)$ and can explicitly be expressed as
\begin{equation}
\begin{split}
\tilde h(k,t)&=\frac{1}{2}[h_1(k)+J^{-2}h'_1(k)h_1(k)h'_1(k)\\
&+(h_1(k)-J^{-2}h'_1(k)h_1(k)h'_1(k))\cos(2Jt)\\
&+iJ^{-1}[h_1(k),h'_1(k)]\sin(2Jt)].
\end{split}
\label{thktf}
\end{equation}
In the following, we will focus on the K-theory classification of $\tilde h(k,t)$ in ($1+1$) dimensions. The stable topology should be inherited by those parent Hamiltonians (\ref{Ht}) that are continuously connected to $\hat H(t)=\sum_k\hat{\boldsymbol{c}}^\dag_k\tilde h(k,t)\hat{\boldsymbol{c}}_k$ but with neither spatial nor temporal periodicity.

\subsection{Symmetry constraints and the maximal K-groups}
Let us discuss the symmetry properties of $\tilde h(k,t)$. Since $h_1(k)$ and $h'_1(k)$ belong to the same AZ class, we have 
\begin{equation}
\begin{split}
\mathcal{A}h_1(k)\mathcal{A}^{-1}=\eta_{\mathcal{A}}h_1(-k),\\
\mathcal{A}h'_1(k)\mathcal{A}^{-1}=\eta_{\mathcal{A}}h'_1(-k),
\end{split}
\end{equation}
where $\eta_{\mathcal{A}}=\pm1$, 
$\mathcal{A}z=z^*\mathcal{A}$ for all $z\in\mathbb{C}$ and $\mathcal{A}^2=\pm1$. Noting the anti-unitary nature of $\mathcal{A}$, we obtain
\begin{equation}
\mathcal{A}\tilde h(k,t)\mathcal{A}^{-1}=\eta_{\mathcal{A}}\tilde h(-k,-\eta_{\mathcal{A}}t).
\label{aham}
\end{equation}
This result can be contrasted to the case of unitary (anti-)symmetries:
\begin{equation}
\begin{split}
Uh_1(k)U^{-1}=\eta_Uh_1(\epsilon_Uk),\\
Uh'_1(k)U^{-1}=\eta_Uh'_1(\epsilon_Uk),
\end{split}
\end{equation}
with $\eta_U=\pm1$ and $\epsilon_U=\pm 1$, leading to
\begin{equation}
U\tilde h(k,t)U^{-1}=\eta_U\tilde h(\epsilon_Uk,\eta_Ut).
\label{uhu}
\end{equation}

\begin{table*}[tbp]
\caption{Topological classification of $\tilde h(k,t)$ subject to the symmetry constraints given in Eq.~(\ref{aham}) (or Eq.~(\ref{uhu}) for class AIII) with $\mathcal{A}=\mathcal{T}$ or/and $\mathcal{C}$. The last column shows a subset of the maximal K-group that can be realized in quench dynamics, i.e., in the form of Eq.~(\ref{flatPBH}). 
The case of class BDI and class D (marked in red) has been discussed in the main text. Other nontrivial classes not appearing in the main text are marked in blue.}
\begin{center}
\begin{tabular}{ccccccc}
\hline\hline
AZ class 
& \;TRS\; & \;PHS\; & \;CS\; & \;\;\;\;\;\;\;\;\;\;\;\;\;\;\;\;\;\;\;\;\;Symmetry constraints on $\tilde h(k,t)$\;\;\;\;\;\;\;\;\;\;\;\;\;\;\;\;\;\;\;\;\; & \;Maximal K-group\; & Dynamical realization \\
\hline
A & 0 & 0 & 0 & None & $\mathbb{Z}$ & 0 \\
AIII & 0 & 0 & 1 & $\Gamma\tilde h(k,t)\Gamma^{-1}=-\tilde h(k,-t)$ & $\mathbb{Z}\oplus\mathbb{Z}$ & \textcolor{blue}{$\mathbb{Z}$} \\
\hline
AI & 1 & 0 & 0 & $\mathcal{T}\tilde h(k,t)\mathcal{T}^{-1}=\tilde h(-k,-t)$ & 0 & 0 \\
BDI & 1 & 1 & 1 & $\mathcal{T}\tilde h(k,t)\mathcal{T}^{-1}=\tilde h(-k,-t)$,\;\;\;\;$\mathcal{C}\tilde h(k,t)\mathcal{C}^{-1}=-\tilde h(-k,t)$ & $\mathbb{Z}$ & \textcolor{red}{$\mathbb{Z}$} \\
D & 0 & 1 & 0 & $\mathcal{C}\tilde h(k,t)\mathcal{C}^{-1}=-\tilde h(-k,t)$ & $\mathbb{Z}_2$ & \textcolor{red}{$\mathbb{Z}_2$} \\
DIII & $-1$ & 1 & 1 & $\mathcal{T}\tilde h(k,t)\mathcal{T}^{-1}=\tilde h(-k,-t)$,\;\;\;\;$\mathcal{C}\tilde h(k,t)\mathcal{C}^{-1}=-\tilde h(-k,t)$ & $\mathbb{Z}_2\oplus\mathbb{Z}_2$ & \textcolor{blue}{$\mathbb{Z}_2$} 
\\
AII & $-1$ & 0 & 0 & $\mathcal{T}\tilde h(k,t)\mathcal{T}^{-1}=\tilde h(-k,-t)$ & $\mathbb{Z}_2$ & 0 \\
CII & $-1$ & $-1$ & 1 & $\mathcal{T}\tilde h(k,t)\mathcal{T}^{-1}=\tilde h(-k,-t)$,\;\;\;\;$\mathcal{C}\tilde h(k,t)\mathcal{C}^{-1}=-\tilde h(-k,t)$ & $\mathbb{Z}$ & \textcolor{blue}{$\mathbb{Z}$} \\
C & 0 & $-1$ & 0 & $\mathcal{C}\tilde h(k,t)\mathcal{C}^{-1}=-\tilde h(-k,t)$ & 0 & 0 \\
CI & 1 & $-1$ & 1 & $\mathcal{T}\tilde h(k,t)\mathcal{T}^{-1}=\tilde h(-k,-t)$,\;\;\;\;$\mathcal{C}\tilde h(k,t)\mathcal{C}^{-1}=-\tilde h(-k,t)$ & 0 & 0 \\
\hline\hline
\end{tabular}
\end{center}
\label{tableS1}
\end{table*}

We first perform a complete classification on the basis of the above symmetry constraints alone. Without any symmetry requirement, $\tilde h(k,t)$ simply belongs to class A in two dimensions and is characterized by $\mathbb{Z}$.

We now turn to the case of TRS alone, i.e., $\mathcal{A}=\mathcal{T}$ ($\eta_{\mathcal{T}}=1$). In this case, we have
\begin{equation}
\mathcal{T}\tilde h(k,t)\mathcal{T}^{-1}=\tilde h(-k,-t),
\label{THT}
\end{equation}
which turns out to be the standard symmetry constraints of classes AI or AII in two dimensions. Therefore, the classification is $0$ (trivial) for $\mathcal{T}^2=1$ (class AI), and is 
$\mathbb{Z}_2$ for $\mathcal{T}^2=-1$ (class AII). 

We then move on to the case of PHS alone, i.e., $\mathcal{A}=\mathcal{C}$ ($\eta_{\mathcal{C}}=-1$). The symmetry constraint on $\tilde h(k,t)$ reads
\begin{equation}
\mathcal{C}\tilde h(k,t)\mathcal{C}^{-1}=-\tilde h(-k,t),
\end{equation}
which coincides with that of 
an adiabatic fermion-parity pump \cite{Teo2010}. The effective dimension is $1-1=0$, so that the classification is $0$ for $\mathcal{C}^2=-1$ (class C), and is $\mathbb{Z}_2$ for $\mathcal{C}^2=1$ (class D). 

Now let us consider the case in which there are both TRS and PHS. Recalling Eq.~(\ref{THT}), we can treat this case as if we add an anti-unitary anti-symmetry with $d_\parallel=1$ (the number of momentum components that do not flip under the symmetry operation) into class AI or AII. We can apply the formula developed in Ref.~\cite{Shiozaki2014}:
\begin{equation}
K^{U/A}_{\mathbb{R}}(s,t;d,d_{\parallel})=K^{U/A}_{\mathbb{R}}(s-d,t-d_{\parallel};0,0),
\end{equation}
where $d=2$, $d_{\parallel}=1$, and $s$ and $t$ are determined by the base system (real AZ class, here it is either class AI or AII) and the additional two-fold symmetries. For class BDI, we have $s=0$ and $t=3$, leading to $\pi_0({\rm C}_{-2})=\mathbb{Z}$. For class CI, we have $s=0$ and $t=1$, leading to $\pi_0({\rm R}_6\times{\rm R}_6)=0$. For class DIII, we have $s=4$ and $t=1$, leading to $\pi_0({\rm R}_2\times{\rm R}_2)=\mathbb{Z}_2\oplus\mathbb{Z}_2$. For class CII, we have $s=4$ and $t=3$, leading to $\pi_0({\rm C}_2)=\mathbb{Z}$. Here, $\pi_0$ is the zeroth homotopy group and ${\rm C}_s={\rm C}_{s+2}$ (${\rm R}_s={\rm R}_{s+8}$) denotes the complex (real) Clifford-algebra extension ${\rm C}\ell_s\to{\rm C}\ell_{s+1}$ (${\rm C}\ell_{0,s}\to{\rm C}\ell_{0,s+1}$) \cite{Kitaev2009}.

For the remaining class AIII, we have to use another formula in Ref.~\cite{Shiozaki2014}:
\begin{equation}
K^U_{\mathbb{C}}(s,t;d,d_{\parallel})=K^U_{\mathbb{C}}(s-d,t-d_{\parallel};0,0),
\label{KUC}
\end{equation}
where $d=2$, $d_{\parallel}=1$, and $s$ and $t$ are determined by the base system (complex AZ class) and the additional two-fold symmetry. Since we start from class A, we have $s=0,t=1$, leading to $\pi_0({\rm C}_{-2}\times{\rm C}_{-2})=\mathbb{Z}\oplus\mathbb{Z}$.

We summarize all the results in the second rightmost column in Table~\ref{tableS1}. Remarkably, all the AZ classes characterized by a trivial maximal K-group turn out to be trivial classes in one dimension \cite{Ryu2008,Kitaev2009,Ryu2010}. The converse is not true, since classes A and AII are trivial in one dimension, whereas the maximal groups are not. It is worthwhile to mention that the formulas developed in Ref.~\cite{Shiozaki2014} are applicable to arbitrary dimensions and arbitrary two-fold-symmetry classes, which can be represented by certain Clifford-algebra extensions. 

Finally, let us discuss the case of a reflection symmetry alone. We can again apply Eq.~(\ref{KUC}), but with $d=2,d_\parallel=0,s=0$ and $t=1$, leading to $\pi_0({\rm C}_{-1})=0$. In fact, such a result was already reported in Ref.~\cite{Ryu2013}. This explains the destruction of ES crossings found in Fig.~\ref{figS5} when we introduce on-site random potential into the SSH model, even if the potential respects the inversion symmetry.

\subsection{Topological numbers and the dynamical realizations}
In this section, we identify all the topological numbers indicated by the K-groups and demonstrate that some of the topological numbers always vanish in quench dynamics. 

\subsubsection{Complex AZ classes}
We start with class A. Since there is no symmetry constraint, the topological number $\mathbb{Z}$ is nothing but the Chern number:
\begin{equation}
C=\iint\frac{dt dk}{16\pi i J^3}
{\rm Tr}[\tilde h(k,t)[\partial_k\tilde h(k,t),\partial_t\tilde h(k,t)]],
\label{Chern}
\end{equation}
Without specifying the form of $\tilde h(k,t)$, the generator of the maximal K-group $\mathbb{Z}$ can be exemplified by a Thouless pump with a unit Chern number.

As for class AIII, the classification $\mathbb{Z}\oplus\mathbb{Z}$ differs significantly from the trivial result ($0$) for the conventional AZ class \cite{Ryu2008}. This is due to the fact that \emph{one} of the quasi-momentum components changes its sign upon being acted on by the chiral operator $\Gamma$. The Chern number can thus be nonzero:
\begin{equation}
\begin{split}
C&\propto i\iint dtdk{\rm Tr}[\tilde h(k,t)[\partial_k\tilde h(k,t),\partial_t\tilde h(k,t)]]\\
&=i\iint dtdk{\rm Tr}[\Gamma\tilde h(k,t)\Gamma[\partial_k\Gamma\tilde h(k,t)\Gamma,\partial_t \Gamma \tilde h(k,t)\Gamma]]\\
&=-i\iint dtdk{\rm Tr}[\tilde h(k,-t)[\partial_k\tilde h(k,-t),\partial_t\tilde h(k,-t)]]\\
&=i\iint d\bar t dk{\rm Tr}[\tilde h(k,\bar t)[\partial_k\tilde h(k,\bar t),\partial_{\bar t}\tilde h(k,\bar t)]].
\end{split}
\end{equation}
This is to be contrasted with the conventional class AIII, where both quasi-momenta are not reversed so that the chiral symmetry enforces the Chern number to be zero. 

The other topological number can be identified as follows. Due to the reversion of one of the quasi-momenta, the chiral symmetry is similar to a reflection (crystalline) symmetry and determines two high symmetry points $t=0,\frac{\pi}{2J}$, where the Hamiltonian $h(k,t)$ is chirally symmetric, i.e., 
\begin{equation}
\begin{split}
\{\tilde h(k,0),\Gamma\}=0,\;\;\;\;\left\{\tilde h\left(k,\frac{\pi}{2J}\right),\Gamma\right\}=0.
\end{split}
\end{equation}
While we can define two winding numbers for these two Hamiltonians, only the \emph{difference} $\Delta W$ between the two winding numbers at high symmetry points is a genuine two-dimensional topological number \cite{Ryu2013,Shiozaki2014}. Indeed, a system with $\Delta W=0$ can be created by stacking a one-dimensional chain along the time direction and thus the system does not possess a nontrivial \emph{two-dimensional} topology. In fact, we have $\frac{1}{2}(\Delta W+C)\in\mathbb{Z}$, due to the quantization of the Chern number over \emph{half} of the momentum-time space with the boundaries at $t=0$ and $\frac{\pi}{2J}$ contracted (leading to a boundary contribution $\frac{1}{2}\Delta W$). The two generators of the K-group can thus be exemplified by the Rice-Mele-Thouless pump \cite{Monika2016b,Takahashi2016} ($C=1,\Delta W=1$) and the quench dynamics in the SSH model across the phase boundary ($C=0,\Delta W=2$).

So far we only impose the symmetry requirement and have not yet specified the form of the Bloch Hamiltonian. If we confine ourselves to the quench dynamics, the flattened parent Hamiltonian takes the form of Eq.~(\ref{flatPBH}). The integrand in Eq.~(\ref{Chern}) can thus be calculated explicitly, giving
\begin{equation}
\begin{split}
&i{\rm Tr}[\tilde h(k,t)[\partial_k\tilde h(k,t),\partial_t\tilde h(k,t)]]\\
=&{\rm Tr}[h_1(k)[[A(k,t),h_1(k)]+\partial_k h_1(k),[h'_1(k),h_1(k)]]]\\
=&{\rm Tr}[([A(k,t),h_1(k)]+\partial_k h_1(k))[[h'_1(k),h_1(k)],h_1(k)]]\\
=&2{\rm Tr}[[A(k,t),h_1(k)](h'_1(k)-h_1(k)h'_1(k)h_1(k))]\\
&+2{\rm Tr}[\partial_k h_1(k)(h'_1(k)-h_1(k)h'_1(k)h_1(k))]\\
=&4J^2({\rm Tr}[A(k,t)[h_1(k),h'_1(k)]]+{\rm Tr}[h'_1(k)\partial_k h_1(k) ]),
\end{split}
\label{intg}
\end{equation}
where we have used $h_1(k)\partial_k h_1(k)+(\partial_k h_1(k))h_1(k)=0$ and
\begin{equation}
\begin{split}
\partial_k\tilde h(k,t)&=e^{-ih'_1(k)t}[A(k,t),h_1(k)]e^{ih'_1(k)t}\\
&+e^{-ih'_1(k)t}\partial_k h_1(k)e^{ih'_1(k)t}
\end{split}
\end{equation}
with $A(k,t)=e^{ih'_1(k)t}\partial_k(e^{-ih'_1(k)t})$, and
\begin{equation}
\partial_t \tilde h(k,t)=-ie^{-ih'_1(k)t}[h'_1(k),h_1(k)]e^{ih'_1(k)t}.
\end{equation}
Noting that $h'^2_1(k)=J^2$, $A(k,t)$ can be expressed as
\begin{equation}
A(k,t)=\left[\frac{\sin^2 Jt}{J^2} h'_1(k)-i\frac{\sin (2Jt)}{2J}\right]\partial_k h'_1(k),
\end{equation}
and its time integral gives 
\begin{equation}
\int^{\frac{\pi}{J}}_0dt A(k,t)=\frac{\pi}{2J^3}h'_1(k)\partial_k h'_1(k).
\label{expA}
\end{equation}
Combining Eqs.~(\ref{intg}) and (\ref{expA}), we obtain
\begin{equation}
C=-\frac{1}{4J^2}\int dk\partial_k{\rm Tr}[h'_1(k)h_1(k)],
\label{Csim}
\end{equation}
which vanishes if the integral range runs over the entire Brillouin zone. 

As for parent Hamiltonians generated by pre- and postquench Hamiltonians 
in class AIII, the Chern number must vanish as well. On the other hand, $\Delta W$ can be nonzero, so a subset $\{(C,\Delta W):C=0,\Delta W\in2\mathbb{Z}\}$, which is isomorphic to $\mathbb{Z}$, can be realized. Furthermore, we have
\begin{equation}
\Delta W
=2(w'-w),
\label{wmw}
\end{equation}
where $w'$ ($w$) is the winding numbers of $h'_1(k)$ ($h_1(k)$). 

Let us show Eq.~(\ref{wmw}) as follows. According to Eq.~(\ref{flatPBH}), the Bloch Hamiltonians at hight symmetry points read
\begin{equation}
\tilde h(k,0)=h_1(k),\;\;\;\;
\tilde h\left(k,\frac{\pi}{2J}\right)=h'_1(k)h_1(k)h'_1(k),
\end{equation}
where, under the choice of $\Gamma=\sigma^z$, $h_1(k)$ and $h'_1(k)$ take the forms
\begin{equation}
h_1(k)=\begin{bmatrix} 0 & u(k) \\ u^\dag(k) & 0 \end{bmatrix},\;\;\;\;
h'_1(k)=\begin{bmatrix} 0 & u'(k) \\ u'^\dag(k) & 0 \end{bmatrix}.
\end{equation}
Therefore, we have
\begin{equation}
h'_1h_1h'_1=\begin{bmatrix} 0 & u'u^\dag u' \\ u'^\dag uu'^\dag & 0 \end{bmatrix},
\end{equation}
implying $w_{t=\frac{\pi}{2J}}=2w'-w$, where the winding numbers are given by $w=\int^\pi_{-\pi}\frac{dk}{2\pi i}\partial_k\ln{\rm det}u(k)$ and $w'=\int^\pi_{-\pi}\frac{dk}{2\pi i}\partial_k\ln{\rm det}u'(k)$. Since $w_{t=0}=w$, we finally obtain Eq.~(\ref{wmw}).

\subsubsection{Real AZ classes}
According to the K-theory, there are five possibly nontrivial (with nontrivial maximal K-group) real AZ classes: BDI, D, DIII, AII and CII (see Table~\ref{tableS1}). We discuss them one by one.

The simplest case is  class AII, of which the maximal K-group is exactly the same as the conventional result $\mathbb{Z}_2$. According to the seminal work by Moore and Balents \cite{Moore2007}, such a $\mathbb{Z}_2$ can be determined from the parity (odd or even) of the Chern number of \emph{half} of the Brillouin zone $[0,\pi]\times[0,\frac{\pi}{J}]$ (a cylinder) after contracting the boundaries while keeping TRS. This is always possible because class AII is trivial in one dimension \cite{Ryu2008}. As for quench dynamics, since class AII is trivial in zero dimension \cite{Ryu2010}, we can always find two paths parameterized by $\gamma\in[0,1]$, which satisfies 
\begin{equation}
h'_{\rm L}(\gamma=0)=h_1(0),\;\;\;\;
h'_{\rm L}(\gamma=1)=h'_1(0),
\end{equation}
and
\begin{equation}
h'_{\rm R}(\gamma=0)=h'_1(\pi),\;\;\;\;
h'_{\rm R}(\gamma=1)=h_1(\pi),
\end{equation}
respectively. The parent Hamiltonians are thus given by
\begin{equation}
\tilde h_{\rm L,R}(\gamma,t)=e^{-ih'_{\rm L,R}(\gamma)t}h_1(k_{\rm L,R})e^{ih'_{\rm L,R}(\gamma)t},
\end{equation}
where $k_{\rm L}=0$ and $k_{\rm R}=\pi$. 
Using Eq.~(\ref{Csim}), we have
\begin{equation}
C=\frac{1}{4J^2}({\rm Tr}[h^2_1(0)]-{\rm Tr}[h^2_1(\pi)])=0,
\end{equation}
so the $\mathbb{Z}_2$ index for a quench dynamics generated by two class AII Hamiltonians is always trivial. On the other hand, with the symmetry constraint alone, the maximal 
K-group can be generated by the Fu-Kane pump \cite{Fu2006}, which is built upon two copies of the Rice-Mele-Thouless pump with opposite spins. Such a $\mathbb{Z}_2$ index stays well-defined in the presence of spin-orbit coupling terms that respect TRS.

\begin{figure}
\begin{center}
        \includegraphics[width=8cm, clip]{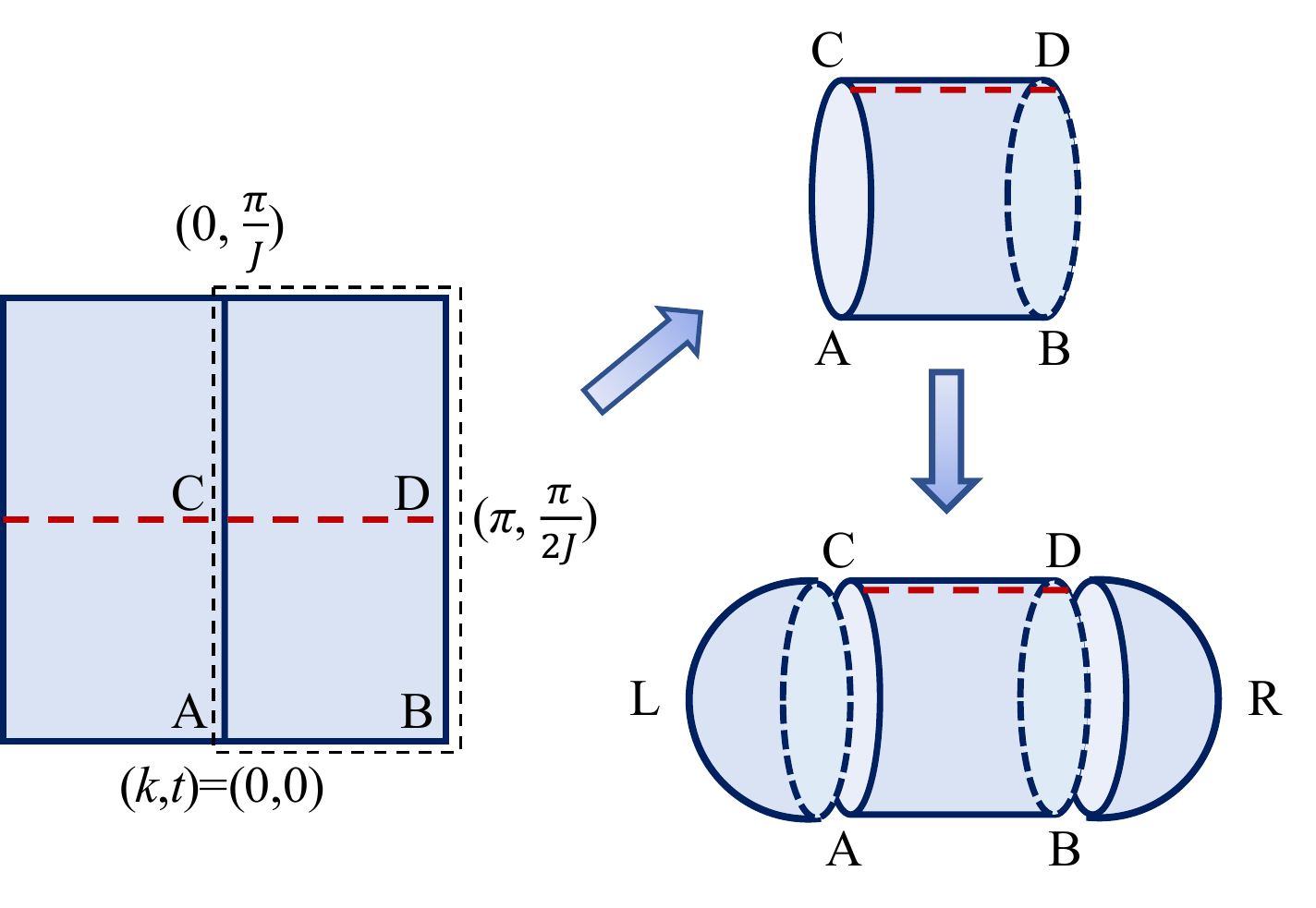}
      \end{center}
   \caption{Moore-Balents approach for calculating the $\mathbb{Z}_2$ index \cite{Moore2007}. The right half (delimited by the dashed rectangle) of a two-dimensional Brillouin zone is equivalent to a cylinder. By contracting the two boundaries of the cylinder while keeping the symmetry (e.g., TRS), we can compactify the manifold, on which a Chern number is now well-defined. Provided that the ambiguity of contraction leads to an even-integer difference, the parity (even or odd) of the Chern number should be a well-defined $\mathbb{Z}_2$ topological index.}
   \label{figS1}
\end{figure}

We move on to class D, which is similar to class AII due to the same $\mathbb{Z}_2$ characterization. Indeed, we can again use the Moore-Balents approach, since the Bloch Hamiltonians at boundaries ($k=0,\pi$) are classified by 0 due to the fact that the effective dimension is $\delta=0-1=-1$, in which class D is trivial \cite{Teo2010}. As for quench dynamics, however, a matrix (zero-dimensional Hamiltonian) in class D is characterized by $\pi_0({\rm R}_2)=\mathbb{Z}_2$, so it is not always possible to deform $h'_1(\Gamma)$ into $h_1(\Gamma)$. Nevertheless, we can always deform $h'_1(\Gamma)$ and $h_1(\Gamma)$ to make them \emph{commute} with each other, even if their $\mathbb{Z}_2$ indices are different. That is, under the basis where $\mathcal{C}=\mathcal{K}$ (complex conjugation), we can always deform a $2n\times2n$ class D matrix $h$, which satisfies 
\begin{equation}
h^*=-h,\;\;h^\dag=h\;\Leftrightarrow\;(ih)^*=ih,\;\;(ih)^{\rm T}=-ih,
\label{Dcons}
\end{equation}
into either $\bigoplus^n_{j=1}\sigma^y$ or $(-\sigma^y)\oplus\bigoplus^{n-1}_{j=1}\sigma^y$, depending on the sign of the Pfaffian of $ih$, which is an anti-symmetric real matrix (see Eq.~(\ref{Dcons})). Neglecting the $2\mathbb{Z}$ ambiguity (class D is characterized by $2\mathbb{Z}$ in the effective dimension $\delta=0-2=-2$ \cite{Teo2010}) 
of contraction, using Eq.~(\ref{Csim}), we can explicitly write down the $\mathbb{Z}_2$ index as
\begin{equation}
\begin{split}
v&=\frac{1}{2}(|{\rm sgn\;Pf}[ih'_1(0)]-{\rm sgn\;Pf}[ih_1(0)]|\\
&-|{\rm sgn\;Pf}[ih'_1(\pi)]-{\rm sgn\;Pf}[ih_1(\pi)]|)\;{\rm mod}\;2,\\
&=|\mathcal{N}'-\mathcal{N}|,
\end{split}
\label{PfZ2}
\end{equation}
where ${\rm sgn\;}x\equiv\frac{x}{|x|}$ and $\mathcal{N}=\frac{1}{2}|{\rm sgn\;Pf}[ih_1(\pi)]-{\rm sgn\;Pf}[ih_1(0)]|$ ($\mathcal{N}'=\frac{1}{2}|{\rm sgn\;Pf}[ih'_1(\pi)]-{\rm sgn\;Pf}[ih'_1(0)]|$). It is worth mentioning that this $\mathbb{Z}_2$ index is generally difficult to calculate due to the PHS constraint during the boundary contraction \cite{Teo2010}. Nevertheless, due to the specific form of the parent Hamiltonians for quench dynamics (\ref{flatPBH}), obtaining the explicit expression (\ref{PfZ2}) now becomes possible.

Now we turn to the real AZ classes with chiral symmetries. We first consider class BDI, which can be obtained from class AIII by adding an involutory ($\mathcal{T}^2=1$), commutative ($[\mathcal{T},\Gamma]=0$) TRS. In the presence of TRS, we can show that the Chern number must vanish:
\begin{equation}
\begin{split}
&C\propto i\iint dtdk{\rm Tr}[\tilde h(k,t)[\partial_k\tilde h(k,t),\partial_t\tilde h(k,t)]]\\
&=-i\iint dtdk{\rm Tr}[\mathcal{T}\tilde h\mathcal{T}^{-1}[\partial_k\mathcal{T}\tilde h\mathcal{T}^{-1},\partial_t\mathcal{T}\tilde h\mathcal{T}^{-1}]]\\
&=-i\iint dtdk{\rm Tr}[\tilde h(-k,-t)[\partial_k\tilde h(-k,-t),\partial_t\tilde h(-k,-t)]]\\
&=-i\iint d\bar t d\bar k{\rm Tr}[\tilde h(\bar k,\bar t)[\partial_{\bar k}\tilde h(\bar k,\bar t),\partial_{\bar t}\tilde h(\bar k,\bar t)]].
\end{split}
\end{equation}
Here we have used ${\rm Tr}[\mathcal{T}A\mathcal{T}]={\rm Tr}[A^\dag]$. On the other hand, the difference between the winding number at high-symmetry lines $t=0,\frac{\pi}{2J}$ can still be nonzero, even if the Hamiltonian takes the form of Eq.~(\ref{flatPBH}). Moreover, since the weak $\mathbb{Z}_2$ number conserves along the $t$ direction, $\Delta W$ must be even and we can define an integer
\begin{equation}
\Delta w=\frac{1}{2}\Delta W
\end{equation}
no matter whether or not $h(k,t)$ is generated by quench dynamics. This explains why the maximal K-group and the dynamical realization are both given by $\mathbb{Z}$. 

Using a similar argument, we can explain why the maximal K-group of class CII also reduces from $\mathbb{Z}\oplus\mathbb{Z}$ to $\mathbb{Z}$ when adding an anti-involutory (i.e., $\mathcal{T}^2=-1$), commutative TRS. Just like class BDI, the remaining $\mathbb{Z}$ corresponds to the difference of winding numbers at high symmetry lines, which is twice the winding-number difference between the pre- and postquench Hamiltonians. To realize the generator, we can construct a spin-orbit coupled SSH-like model
\begin{equation}
h(k)=-[(J_1+J_2\cos k)\sigma^z\otimes\sigma^y+J_2\sin k\sigma^x\otimes\sigma^0],
\end{equation}
for which we do the quench
\begin{equation}
(J_1,J_2)=(J,0)\to(0,J).
\end{equation}
Here $\mathcal{T}=i\sigma^y\otimes\sigma^0\mathcal{K}$ and $\mathcal{C}=i\sigma^0\otimes\sigma^y\mathcal{K}$.

Finally, we consider class DIII, which is characterized by $\mathbb{Z}_2\oplus\mathbb{Z}_2$. The simplest way to understand this topological classification is to regard class DIII as class AIII with an additional  anti-involutory and anti-commutative ($\{\mathcal{T},\Gamma\}=0$) TRS. Note that class AIII is characterized by $\mathbb{Z}\oplus\mathbb{Z}$. Due to the additional TRS, both the winding number at high symmetry lines and the Chern number vanish. Nevertheless, it is still
possible to have a nontrivial Chern number or/and a nontrivial winding number over one half of the Brillouin zone after deforming the boundary to compactify the manifold (i.e., using the Moore-Balents approach). While both the Chern number and the winding number are ambiguous, their parities are unique. Therefore, one of the $\mathbb{Z}_2$ index should be the same as that in class AII, while the other equals to the difference of $\mathbb{Z}_2$ index at high symmetry lines, where the one-dimensional section belongs to class DIII. 

If the Hamiltonian is generated by quench dynamics, the TRS-related $\mathbb{Z}_2$ index vanishes, as we have proved for class AII. On the other hand, the other $\mathbb{Z}_2$ index could be nonzero. To realize the generator, we can construct an explicit model --- the spin-$\frac{1}{2}$ SSH model 
\begin{equation}
h(k)=-\sigma^0\otimes[(J_1+J_2\cos k)\sigma^x+J_2\sin k\sigma^y],
\label{halfSSH}
\end{equation}
which undergoes the quench
\begin{equation}
(J_1,J_2)=(J,0)\to(0,J).
\end{equation}
Here $\mathcal{T}=i\sigma^y\otimes\sigma^0\mathcal{K}$ and $\mathcal{C}=\sigma^0\otimes\sigma^z\mathcal{K}$. Note that the hopping in Eq.~(\ref{halfSSH}) does not flip the spin, it is obvious that the TRS is respected. The PHS inherits simply from that of the spinless SSH model. While the other $\mathbb{Z}_2$ index vanishes in quench dynamics, we can make it nontrivial in an adiabatic pump like \cite{Fu2007b}
\begin{equation}
\begin{split}
h(k,t)=&-J_0[1-\cos k-\cos(2Jt)]\sigma^0\otimes\sigma^x\\
&-J_0[\sin(2Jt)\sigma^0\otimes\sigma^y+\sin k\sigma^x\otimes\sigma^z],
\end{split}
\end{equation}
which can be examined to respect both TRS and PHS. 

So far we have identified all the elements in the maximal K-group that are realizable in quench dynamics. As summarized in Table~\ref{tableS1}, the results turn out to be consistent with the classification of topological insulators and superconductors in one dimension \cite{Ryu2008,Ryu2010}. However, we should again emphasize that the topology underlying the quench dynamics in one dimension is of two-dimensional nature. It is also worth mentioning an intuitive understanding on why the quench dynamics in trivial AZ classes must be trivial, even if the maximal K-group is not --- we can always continuously deform $h'_1(k)$ into $h_1(k)$, so that the parent Bloch Hamiltonian $\tilde h(k,t)=h_1(k)$ has no $t$ dependence and thus cannot exhibit genuine two-dimensional nontrivial topology. This argument should also be applicable to higher dimensions and other symmetry classes.

\section{Quench dynamics in two-band models}
In this section, we provide a detailed analysis on the quench dynamics in two-band systems. We focus especially on the SSH model, which has strong experimental relevance. 

\subsection{Dynamics of pesudospins in momentum space}
For general two-band systems such as a superlattice, the pre- and post-quench Bloch Hamiltonians can be expressed as $h(k)=d_0(k)+\boldsymbol{d}(k)\cdot\boldsymbol{\sigma}$ and $h'(k)=d'_0(k)+\boldsymbol{d}'(k)\cdot\boldsymbol{\sigma}$. Denoting the parent Bloch Hamiltonian as
$h(k,t)=d_0(k,t)+\boldsymbol{d}(k,t)\cdot\boldsymbol{\sigma}$ and using Eq.~(\ref{Block}) and the commutation relations for the Pauli matrices, we obtain $d_0(k,t)=d_0(k)$ and
\begin{equation}
\begin{split}
i\partial_td_\kappa(k,t)\sigma^\kappa&=
[d'_\mu(k)\sigma^\mu,d_\nu(k,t)\sigma^\nu]\\
&=2i\epsilon_{\mu\nu\kappa}d'_\mu(k) d_\nu(k,t)\sigma^\kappa,
\end{split}
\label{Eind}
\end{equation}
where $\mu,\nu,\kappa\in\{x,y,z\}$ and the Einstein summation convention is assumed. Equation (\ref{Eind}) can be rewritten as
\begin{equation}
\partial_t\boldsymbol{d}(k,t)=2\boldsymbol{d}'(k)\times\boldsymbol{d}(k,t),
\label{precession}
\end{equation}
Noting that $d_0(k,t)$ plays no role in either dynamics or band topology, we assume it to vanish for simplicity. Indeed, $d_0(k,t)=0$ in both the Su-Schrieffer-Heeger (SSH) model and the Rice-Mele model.

Equation (\ref{precession}) describes the precession of $\boldsymbol{d}(k,t)$ with respect to $\boldsymbol{d}'(k)$ by an angular velocity $2d'(k)$ ($d'(k)\equiv|\boldsymbol{d}'(k)|$). Thus, we can write down the following solution:
\begin{equation}
\begin{split}
\boldsymbol{d}(k,t)=\boldsymbol{d}_{\|}(k)&+\cos(2d'(k)t)\boldsymbol{d}_{\perp}(k)\\
&+\sin(2d'(k)t)\boldsymbol{d}_{\rm o}(k),
\end{split}
\label{genfor}
\end{equation}
where the parallel ($\l$), perpendicular ($\perp$) and out-of-plane ($\rm o$) (spanned by $\boldsymbol{d}(k)$ and $\boldsymbol{d}'(k)$) components are given by
\begin{equation}
\begin{split}
\boldsymbol{d}_{\|}(k)&=[\boldsymbol{d}(k)\cdot\boldsymbol{n}'(k)]\boldsymbol{n}'(k),\\
\boldsymbol{d}_{\perp}(k)&=\boldsymbol{d}(k)-\boldsymbol{d}_{\|}(k),\\
\boldsymbol{d}_{\rm o}(k)&=\boldsymbol{d}(k)\times\boldsymbol{n}'(k),
\end{split}
\label{ppero}
\end{equation}
where $\boldsymbol{n}'(k)\equiv-\frac{\boldsymbol{d}'(k)}{d'(k)}$.

As for the quench in the SSH model, we have
\begin{equation}
\boldsymbol{d}(k)=-(J,0,0),\;\;
\boldsymbol{d}'(k)=-(J'+J\cos k,J\sin k,0).
\label{sshquench}
\end{equation}
Substituting Eq.~(\ref{sshquench}) into Eqs.~(\ref{genfor}) and (\ref{ppero}) yields
\begin{equation}
\begin{split}
d_x(k,t)&=-J+2J\left[\frac{J\sin k\sin(d'(k)t)}{d'(k)}\right]^2,\\
d_y(k,t)&=-2(J'+J\cos k)\sin k\left[\frac{J\sin(d'(k)t)}{d'(k)}\right]^2,\\
d_z(k,t)&=-\frac{J^2}{d(k)}\sin k\sin(2d'(k)t),
\end{split}
\label{sshdkt}
\end{equation}
where 
\begin{equation}
d'(k)=\sqrt{J^2+J'^2+2J'J\cos k}. 
\label{dp}
\end{equation}
Note that $d_z(k,t)$ is generally nonzero, though $d_z(k)=d'_z(k)=0$. We can then obtain the pseudospin texture
\begin{equation}
\boldsymbol{n}(k,t)\equiv-\frac{\boldsymbol{d}(k,t)}{|\boldsymbol{d}(k,t)|}=-\frac{\boldsymbol{d}(k,t)}{J},
\end{equation}
from Eq.~(\ref{sshdkt}).  

\begin{figure}
\begin{center}
        \includegraphics[width=8.5cm, clip]{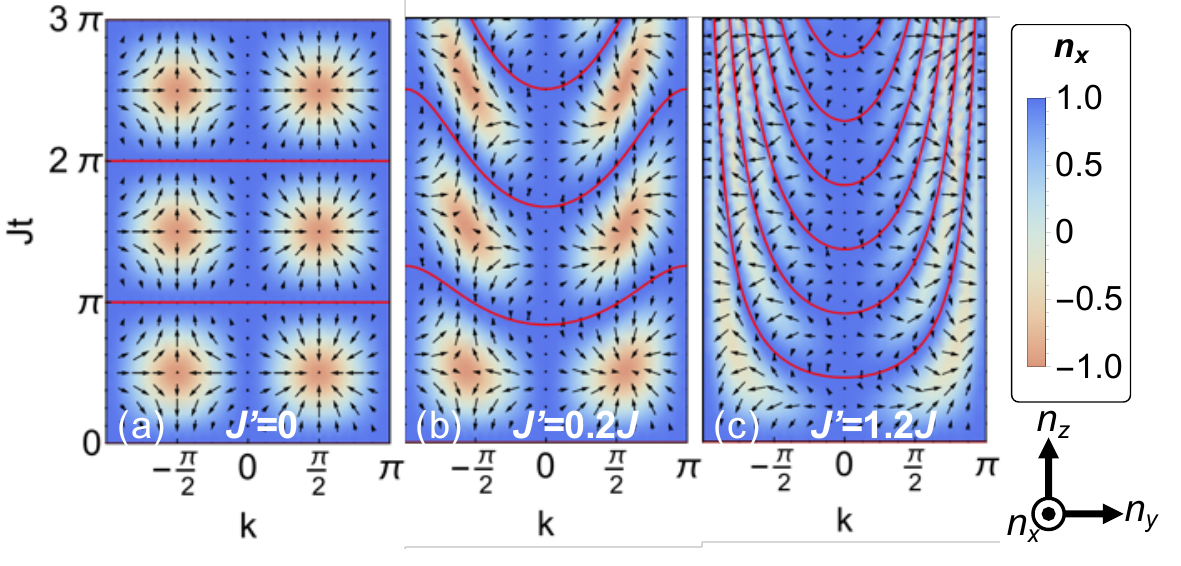}
      \end{center}
   \caption{Pseudospin textures $\boldsymbol{n}(k,t)$ in the $k$-$t$ space for the quench protocols $(J_1,J_2)=(J,0)\to(J',J)$ with (a) $J'=0$, (b) $J'=0.2J$ and (c) $J'=1.2J$. Momentum-time Skyrmions emerge in (a) and (b), but not in (c), as explicitly shown by calculating the topological number defined in Eq.~(\ref{Csky}) (see also Fig. \ref{figS3}). Pseudospins along the red curves are polarized in the $x$ direction.}
   \label{figS2}
\end{figure}

Figure \ref{figS2} shows the dynamics of $\boldsymbol{d}(k,t)$ for the quench in the SSH model. Remarkably, for $J'=0$, the pseudospin texture $\boldsymbol{n}(k,t)$ exhibits rows of 
Skyrmion lattices with opposite topological charges, which can be calculated from 
\begin{equation}
C_{\rm skyrmion}=\iint_{\rm A}\frac{dtdk}{4\pi}
\boldsymbol{n}(k,t)\cdot[\partial_k\boldsymbol{n}(k,t)\times\partial_t\boldsymbol{n}(k,t)],
\label{Csky}
\end{equation}
with ${\rm A}=[0,\frac{\pi}{J}]\times[0,\pi]$ (right column) or $[0,\frac{\pi}{J}]\times[-\pi,0]$ (left). When $0<J'<J$, Skyrmions are still well-defined but deformed and canted. In fact, the pseudospin texture can always be mapped to two arrays of Skyrmion lattices via rescaling the time axis in a $k$-dependent manner, or equivalently via band flattening of the postquench Hamiltonian such that $h'^2(k)=J^2$. Such a continuous deformation breaks down at the critical point $J'=J$, above which $\boldsymbol{n}(k,t)$ becomes topologically trivial since it can continuously be deformed to the pseudospin polarized state. 

\subsection{Dynamical Chern number}
We study in details the Skyrmion charge (\ref{Csky}), which can alternatively be expressed as the integral of the Berry curvature $\Omega(k,t)$ on the $k-t$ space:
\begin{equation}
\Omega(k,t)\equiv2{\rm Im}[(\partial_t\boldsymbol{u}(k,t))^\dag\partial_k\boldsymbol{u}(k,t)].
\label{Okt}
\end{equation}
Here $\boldsymbol{u}(k,t)$ is the lower-band Bloch vector satisfying $h(k,t)\boldsymbol{u}(k,t)=-d(k)\boldsymbol{u}(k,t)$. It should be emphasized that this Berry curvature is merely a mathematical analogy, since the real physical process is a highly nonadiabatic quench dynamics. 

In terms of the eigenvectors of $h'(k)$, which are denoted as $\boldsymbol{v}_\pm(k)$ with eigenvalues $\pm d'(k)$, $\boldsymbol{u}(k,t)$ can explicitly be written as
\begin{equation}
\begin{split}
\boldsymbol{u}(k,t)&=e^{-id'(k)t}\boldsymbol{v}_+(k)(\boldsymbol{v}_+(k))^\dag\boldsymbol{u}(k)\\
&+e^{id'(k)t}\boldsymbol{v}_-(k)(\boldsymbol{v}_-(k))^\dag\boldsymbol{u}(k).
\end{split}
\label{ukt}
\end{equation}
Accordingly, the inner product $(\partial_t\boldsymbol{u}(k,t))^\dag\partial_k\boldsymbol{u}(k,t)$ in Eq.~(\ref{Okt}) is calculated to be
\begin{equation}
\begin{split}
&td'\partial_kd'+id'[\boldsymbol{u}^\dag\boldsymbol{v}_+\partial_k(\boldsymbol{v}_+^\dag\boldsymbol{u})-\boldsymbol{u}^\dag\boldsymbol{v}_-\partial_k(\boldsymbol{v}_-^\dag\boldsymbol{u})]\\
&+id'[|\boldsymbol{u}^\dag\boldsymbol{v}_+|^2(\boldsymbol{v}^\dag_+\partial_k\boldsymbol{v}_+)-|\boldsymbol{u}^\dag\boldsymbol{v}_-|^2(\boldsymbol{v}^\dag_-\partial_k\boldsymbol{v}_-)]\\
&+id'[e^{2id't}\boldsymbol{u}^\dag\boldsymbol{v}_+\boldsymbol{v}^\dag_-\boldsymbol{u}(\boldsymbol{v}^\dag_+\partial_k\boldsymbol{v}_-)+{\rm H.c.}],
\end{split}
\end{equation}
where we have used $\boldsymbol{v}^\dag_+\partial_k\boldsymbol{v}_-+\boldsymbol{v}_-\partial_k\boldsymbol{v}^\dag_+=\partial_k(\boldsymbol{v}^\dag_+\boldsymbol{v}_-)=0$ and temporarily drop the $k$-dependence for simplicity. Since $\boldsymbol{v}^\dag_\pm\partial_k\boldsymbol{v}_\pm$ is purely imaginary and $2{\rm Re}[\boldsymbol{u}^\dag\boldsymbol{v}_\pm\partial_k(\boldsymbol{v}_\pm^\dag\boldsymbol{u})]=\partial_k(|\boldsymbol{u}^\dag(k)\boldsymbol{v}_\pm(k)|^2)$, we have
\begin{equation}
\begin{split}
&\Omega(k,t)=2d'(k)\partial_k(|\boldsymbol{u}^\dag(k)\boldsymbol{v}_+(k)|^2)\\
&-2d'(k)(e^{2id'(k)t}{\rm Tr}[P(k)P_+(k)\partial_kP_+(k)]+{\rm H.c.}),
\end{split}
\label{Oktexpr}
\end{equation}
where $P(k)\equiv\boldsymbol{u}(k)\boldsymbol{u}^\dag(k)$ and $P_+(k)\equiv\boldsymbol{v}_+(k)\boldsymbol{v}^\dag_+(k)$ are projective matrices. 

Note that the second term in Eq.~(\ref{Oktexpr}) oscillates with $t$ due to the factor $e^{2id'(k)t}$. After integrating $\Omega(k,t)$ over ${\rm A}=\{(k,t):0<t<\frac{\pi}{d'(k)},0<k<\pi\}$ in the $k-t$ plane, the contribution from this oscillating term vanishes and we obtain a quantized dynamical Chern number
\begin{equation}
C_{\rm dyn}\equiv\iint_A\frac{dkdt}{2\pi}\Omega(k,t)=F(\pi)-F(0),
\label{Cs}
\end{equation}
where
\begin{equation}
F(k)\equiv|\boldsymbol{u}^\dag(k)\boldsymbol{v}_+(k)|^2.
\end{equation}
The quantization of $C_{\rm dyn}$ is ensured by the particle-hole symmetry (PHS), which restricts the Bloch states of $h(k)$ and $h'(k)$ at $\Gamma=0,\pi$ to be an eigenstate of $\sigma^x$. Concretely, denoting the eigenvalues of $\boldsymbol{u}(\Gamma)$ and $\boldsymbol{v}_-(\Gamma)$ as $\nu_\Gamma$ and $\nu'_\Gamma$, i.e., $\sigma^x\boldsymbol{u}(\Gamma)=\nu_\Gamma\boldsymbol{u}(\Gamma)$ and $\sigma^x\boldsymbol{v}_-(\Gamma)=\nu'_\Gamma\boldsymbol{v}_-(\Gamma)$, we have $F(\Gamma)=\frac{1}{2}|\nu'_\Gamma-\nu_\Gamma|$ and thus
\begin{equation}
C_{\rm dyn}=\frac{1}{2}(|\nu'_\pi-\nu_\pi|-|\nu'_0-\nu_0|).
\label{Cnu}
\end{equation}
This result is consistent with Eq.~(\ref{PfZ2}), which applies to an arbitrary number of bands.

Alternatively, we can introduce
\begin{equation}
C(T)\equiv\int^\pi_0 dk\int^T_0 dt\frac{\Omega(k,t)}{2Td'(k)},
\label{CT}
\end{equation}
and define the dynamical Chern number as
\begin{equation}
C_{\rm dyn}\equiv\lim_{T\to\infty}C(T),
\label{Ci}
\end{equation}
The equivalence between Eqs.~(\ref{Ci}) and (\ref{Cs}) can be understood from the fact that the time integral of the oscillating term in Eq.~(\ref{Oktexpr}) is bounded, and accordingly vanishes after being divided by an infinitely large $T$.

\begin{figure}
\begin{center}
        \includegraphics[width=7cm, clip]{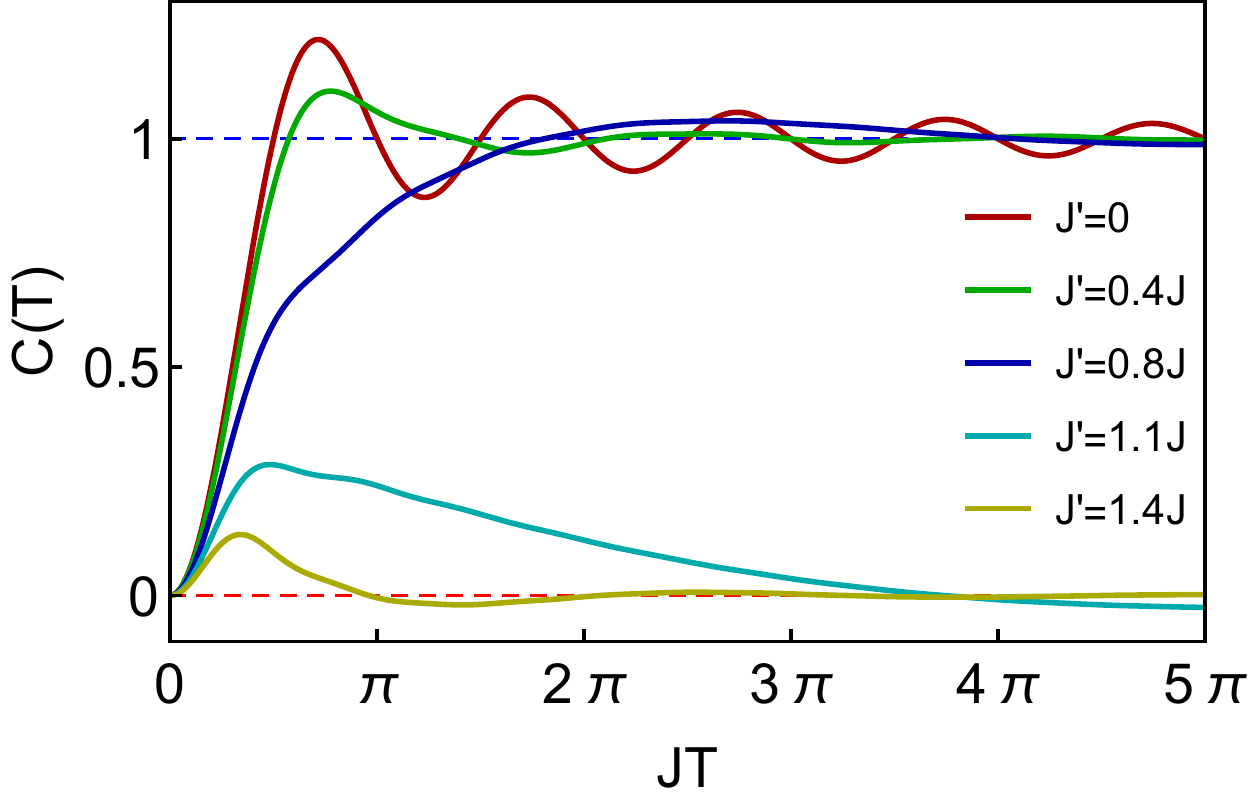}
      \end{center}
   \caption{$C(T)$ defined in Eq.~(\ref{CT}) versus $JT$ in the quench dynamics of the SSH model for different parameters. $C(T)$ converges to $1$ for $J'<J$ (dashed blue line) and $0$ for $J'>J$ (dashed red line).}
   \label{figS3}
\end{figure}

By explicitly calculating $\Omega(k,t)$ for the quench dynamics in the SSH model, we obtain
\begin{equation}
\Omega(k,t)=2\sin k(J+J'\cos k)\left[\frac{J\sin(d'(k)t)}{d'(k)}\right]^2,
\end{equation}
where $d'(k)$ is given in Eq.~(\ref{dp}). Using the definition in Eq.~(\ref{Cs}), we obtain
\begin{equation}
\begin{split}
C_{\rm dyn}&=\int^\pi_0 dk\frac{J^2(J+J'\cos k)\sin k}{2(J^2+J'^2+2J'J\cos k)^{\frac{3}{2}}}\\
&=\int^1_{-1} ds\frac{J^2(J+J's)}{2(J^2+J'^2+2J'Js)^{\frac{3}{2}}}\\
&=\frac{1}{2}[1+{\rm sgn}(J-J')].
\end{split}
\end{equation}
We thus find $C_{\rm dyn}=1$ ($C_{\rm dyn}=0$) if $J>J'$ ($J<J'$), which is consistent with Eq.~(\ref{Cnu}). After straightforward calculations we obtain the expression for $C(T)$ in Eq.~(\ref{CT}):
\begin{equation}
\begin{split}
C(T)&=\frac{1}{2}[1+{\rm sgn}(J-J')]-g\left(T,\frac{J'}{J}\right),\\
g(t,r)&=\int^1_{-1}ds\frac{(1+rs)\sin(2t\sqrt{1+r^2+2rs})}{4t(1+r^2+2rs)^2}.
\end{split}
\end{equation}
We plot $C(T)$ for several different quench parameters in Fig.~\ref{figS3}. It seems that $C(T)$ typically converges more quickly for a finite $J'$ than the flat band case $J'=0$ (but this is not the case when $J'\simeq J$, i.e., close to the critical point). A physical explanation is 
that a finite bandwidth causes certain disorder in the frequency direction and washes out quantum coherent oscillations. We note that similar observations are made in Ref.~\cite{Budich2016}, but in quite a different context of the asymptotic quantization of the nonequilibrium Hall conductance.

\section{Calculation of the entanglement-spectrum dynamics}
In this section, we provide the details on how we obtain the ES dynamics shown in Figs.~\ref{fig1}, \ref{fig2} and \ref{fig3} in the main text. We also provide additional numerical results, especially the many-body-ES dynamics in an interacting system.

\subsection{Numerical method for general noninteracting systems}
We follow the method proposed in Ref.~\cite{Peschel2003} to numerically calculate the single-particle entanglement spectrum (ES). The basic idea is that in a particle-number-conserving free-fermion system described by a quadratic Hamiltonian in terms of $\hat c_j$ ($j=1,2,...,N$), any reduced density operator $\hat\rho_S$ of the ground state $|\Psi\rangle$, which consists of the modes $\hat c_j$ with $j\in S\subset\{1,2,...,N\}$, is a Gaussian state 
\begin{equation}
\hat\rho_S\propto e^{-\sum_{m,n\in S} (h_{\rm E})_{mn}\hat c^\dag_m\hat c_n}\equiv e^{-\hat H_{\rm E}},
\end{equation}
which can be reconstructed from its $|S|\times|S|$ ($|S|$: cardinality of $S$) correlation matrix
\begin{equation}
C_{mn}\equiv{\rm Tr}[\hat c^\dag_m\hat c_n\hat\rho_S]=\langle\Psi|\hat c^\dag_m\hat c_n|\Psi\rangle,\;\;m,n\in S,
\label{corm}
\end{equation}
via 
\begin{equation}
C=\frac{1}{e^{h_{\rm E}}+1}. 
\end{equation}
Therefore, the eigenvalues of $C$ simply gives the single-particle ES \cite{Hughes2011}
\begin{equation}
\xi_n\equiv\frac{1}{e^{\epsilon_n}+1},
\end{equation}
where $\epsilon_n$'s are the eigenvalues of $\hat H_{\rm E}$ or $h_{\rm E}$.

In particular, for a one-dimensional lattice system with $L$ unit cells and subjected to the periodic boundary condition, we can utilize the translational invariance to represent the many-body wave function in a factorized form:
\begin{equation}
|\Psi\rangle=\prod_k\boldsymbol{c}^\dag_k\boldsymbol{u}(k)|{\rm vac}\rangle, 
\label{Psifac}
\end{equation}
where 
$\boldsymbol{u}(k)$ is the Bloch vector and $\boldsymbol{c}^\dag_k\boldsymbol{u}(k)=\sum_\alpha u_\alpha(k)\hat c^\dag_{k\alpha}$ with $\alpha$ being the internal degrees of freedom. In this case, the correlation matrix (\ref{corm}) turns out to be a block-Toeplitz matrix (a matrix $M_{jl}$ is called a Toeplitz matrix if $M_{jl}=M_{j-l}$):
\begin{equation}
\begin{split}
C_{m\alpha,n\beta}&\equiv\langle\Psi|\hat c^\dag_{m\alpha}\hat c_{n\beta}|\Psi\rangle\\
&=\frac{1}{L}\sum_ku^*_\alpha(k)u_\beta(k)e^{ik(m-n)}=C^{\alpha\beta}_{m-n},
\end{split}
\label{Cjalb}
\end{equation}
where $m,n$ are the site indices. To calculate the inter-unit-cell half-chain ES, we only have to figure out the eigenvalues of the $\frac{LD}{2}\times\frac{LD}{2}$ matrix $C_{m\alpha,n\beta}$ (\ref{Cjalb}), where $D$ is the total number of internal degrees of freedom.

\begin{figure}
\begin{center}
        \includegraphics[width=7cm, clip]{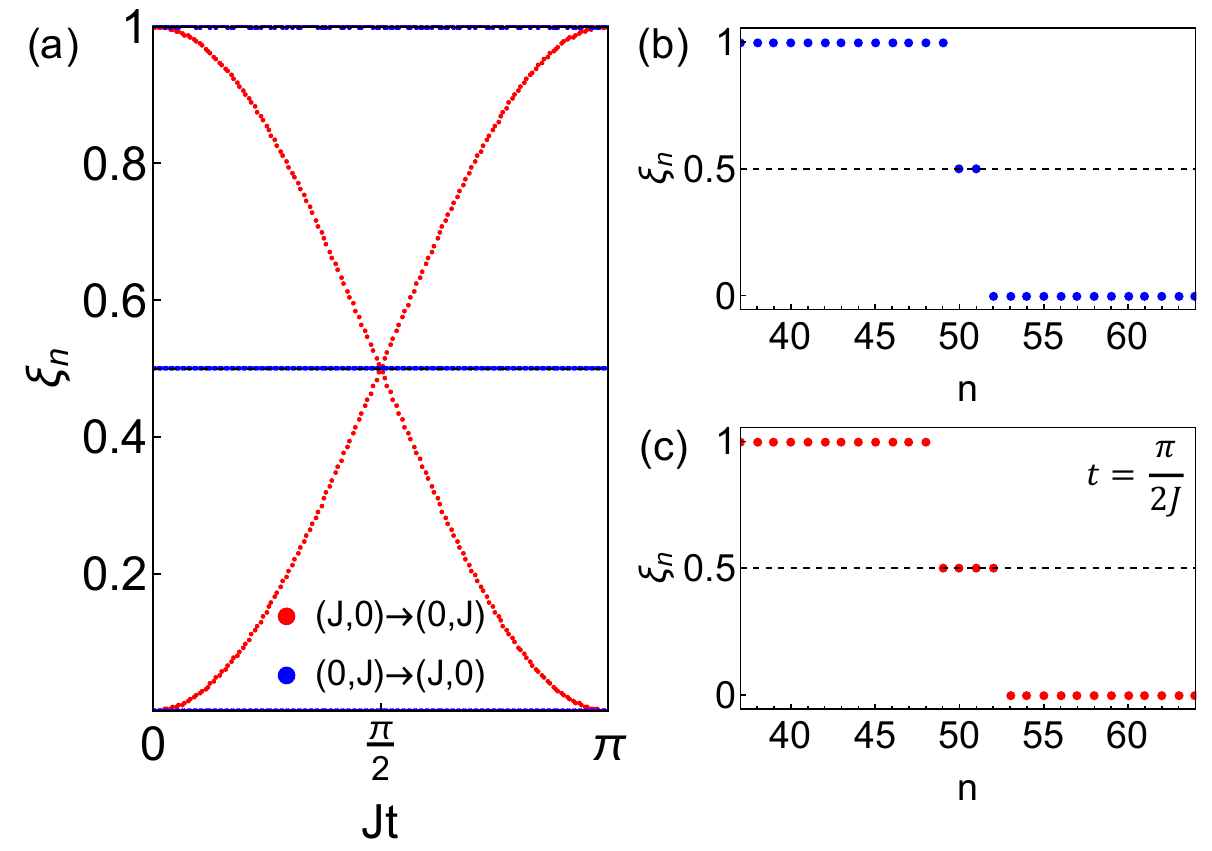}
      \end{center}
   \caption{(a) ES dynamics after the quenches $(J_1,J_2)=(0,J)\to(J,0)$ (blue) and $(J,0)\to(0,J)$ (red) in the SSH model. (b) In the former case, the dynamics is trivial and two degenerate entanglement edge modes at $\xi_n=\frac{1}{2}$ persist. (c) In the latter case, the dynamics is nontrivial, and the instantaneous four-fold degeneracy at $\xi_n=\frac{1}{2}$ emerges at $t=\frac{\pi}{2J}$. 
   }
   \label{figS4}
\end{figure}

In a superlattice system, $D=2$ and $\alpha$ labels the sublattices. By calculating first the dynamics of the Bloch vectors governed by Eq.~(\ref{ukt}) and then 
the correlation matrix (\ref{Cjalb}) followed by exact diagonalization, we can obtain the dynamics of the full single-particle ES. As a benchmark, we plot in Fig.~\ref{figS4}(a) the ES dynamics for the quench $(J_1,J_2)=(0,J)\to(J,0)$ in the SSH model, after which no entanglement is generated for the entanglement cut shown in Fig.~\ref{fig1}(b) in the main text so the ES should stay unchanged. Since the initial state is topologically nontrivial ($\mathcal{N}=2$), we find two degenerate entanglement edge modes at $\xi_n=\frac{1}{2}$, as shown in Fig.~\ref{figS4}(b). This is to be compared with the ES dynamics for the quench $(J,0)\to(0,J)$, after which we find nontrivial dynamics and instantaneous \emph{four}-fold degeneracy at $\xi_n=\frac{1}{2}$ when $t=\frac{\pi}{2J}$ (Fig.~\ref{figS4}(c)).

\begin{figure}
\begin{center}
        \includegraphics[width=8.5cm, clip]{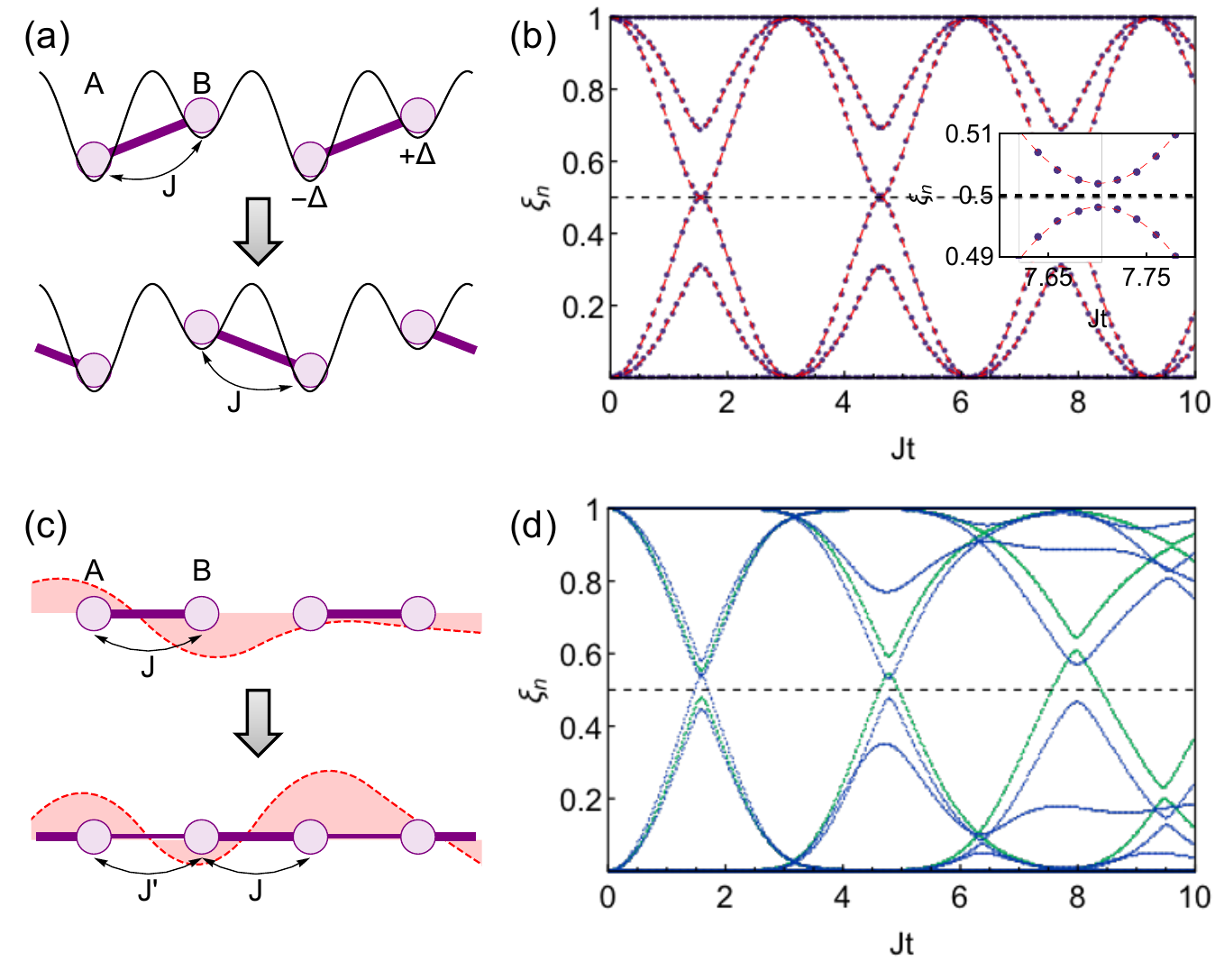}
      \end{center}
   \caption{(a) Quench in the flat-band Rice-Mele model (\ref{HRM}) with $\Delta=0.2J$ and (b) the corresponding ES dynamics. The dots and dashed curves show the numerical and analytical results, respectively. The inset shows the enlarged view around $Jt=7.7$, showing that the two-fold degeneracy is lifted, and crossings are gapped out, although the gap is very small (see inset). (c) Quench from a dimerized state with a random on-site potential $V^{a,b}_j$ shown schematically by a dashed curve (see Eq.~(\ref{RSSH})). Here $W=J'=0.2J$. (d) The corresponding ES dynamics, which no longer exhibit crossings at $\xi_n=\frac{1}{2}$ with (green) and without (blue) inversion symmetry. When $V_j$ respects the inversion symmetry, the ES is two-fold degenerate.}
   \label{figS5}
\end{figure}

According to the notion of symmetry-protected topological phases, we expect that the ES crossings may disappear if the PHS is explicitly broken, for example, by adding a staggered potential $\sum_j\Delta(\hat b^\dag_j\hat b_j-\hat a^\dag_j\hat a_j)$ in the SSH Hamiltonian; then the model becomes the Rice-Mele model \cite{Rice1982} (see Fig.~\ref{figS5}(a)): 
\begin{equation}
\begin{split}
\hat H_{\rm RM}=&-\sum_j(J_1\hat b^\dag_j\hat a_j+J_2\hat a^\dag_{j+1}\hat b_j+{\rm H.c.})\\
&+\sum_j\Delta(\hat b^\dag_j\hat b_j-\hat a^\dag_j\hat a_j).
\end{split}
\label{HRM}
\end{equation}
This is confirmed numerically in the flat-band case, as shown in Fig.~\ref{figS5}(b). An interesting observation here is that the gap is as small as $O(\frac{\Delta^3}{J^3})$, as will be explained in the next subsection. Note that the dynamical Chern number (\ref{Cs}) is no longer well-defined (not quantized) in this case.

We also calculate the ES dynamics in the SSH model with a random on-site potential:
\begin{equation}
\begin{split}
\hat H_{\rm RSSH}=&-\sum_j(J_1\hat b^\dag_j\hat a_j+J_2\hat a^\dag_{j+1}\hat b_j+{\rm H.c.})\\
&+\sum_j(V^a_j\hat a^\dag_j\hat a_j+V^b_j\hat b^\dag_j\hat b_j),
\end{split}
\label{RSSH}
\end{equation}
where $V^a_j$ and $V^b_j$ are randomly sampled from a uniform distribution over $[-W,W]$. We consider the ES dynamics after a quench that changes both the disorder configuration and the following parameters (see Fig.~\ref{figS5}(c)):
\begin{equation}
(J_1,J_2,\Delta)=(J,0,\Delta)\to(J',J,0).
\label{quenRM}
\end{equation}
As shown in Fig.~\ref{figS5}(d), we find that ES crossings at $\xi_n=\frac{1}{2}$ disappear, even if the disorder respects the inversion symmetry, i.e., $V^a_j=V^b_{(\frac{L}{2}-j)\;{\rm mod}\;L+1}$. On the other hand, we will see in Fig.~\ref{figS9} that the ES crossings are robust against disorder in hopping amplitudes, which preserves PHS (and also TRS).

\begin{figure}
\begin{center}
        \includegraphics[width=8.5cm, clip]{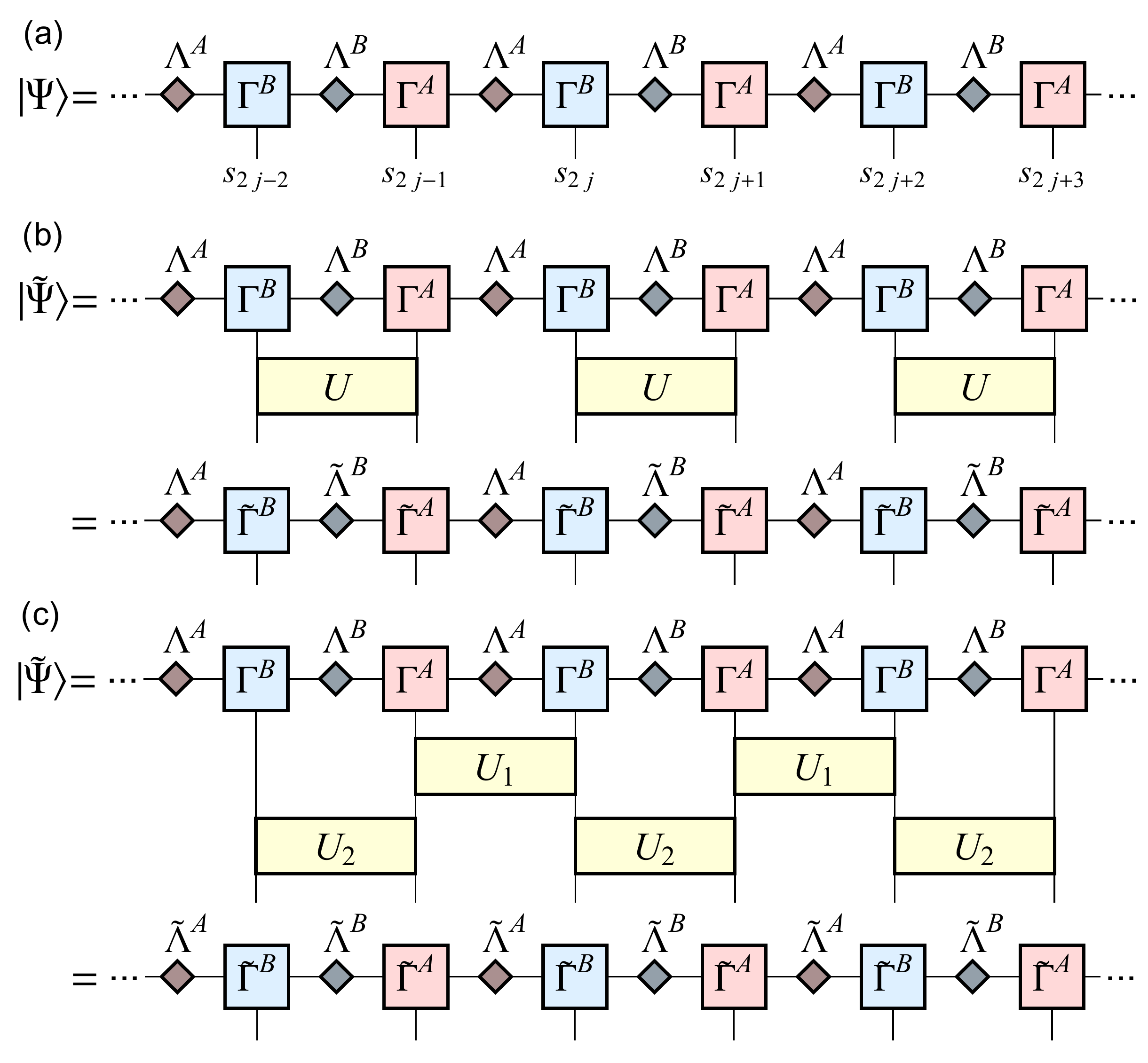}
      \end{center}
   \caption{(a) MPS representation of a one-dimensional superlattice system in the spin language. (b) After a 
   factorized unitary evolution $\hat U=\bigotimes_j\hat U_j$ with translational invariance ($\langle s_{2j}s_{2j+1}|\hat U_j|s'_{2j}s'_{2j+1}\rangle=U^{s_{2j}s_{2j+1}}_{s'_{2j}s'_{2j+1}}$ for $\forall j$, where $s,s'\in\{\uparrow,\downarrow\}$), $|\tilde\Psi\rangle=\hat U|\Psi\rangle$ has a different (the same) ES encoded in $\tilde\Lambda^B$ ($\Lambda^A$) for the intra-unit-cell (inter-unit-cell) entanglement cut. (c) A single iteration in the iTEBD algorithm \cite{Vidal2007}. A general time-evolution operator $\hat U(\delta t)$ can be approximated as $(\bigotimes_j\hat U_{2j})(\bigotimes_j\hat U_{1j})$ via the Suzuki-Trotter decomposition. Unlike the specific case in (b), both $\Lambda^A$ and $\Lambda^B$ are updated.}
   \label{figS6}
\end{figure}

\subsection{Analytical results for some flat-band quenches}
Since the entanglement cut is made in real space, a straightforward way to readout the ES is to represent the real-space many-body wave function in the form of matrix-product state (MPS) \cite{Vidal2007,Pollmann2010}. To do this, we should first translate the picture of a one-dimensional fermionic superlattice into that of a spin chain via the Jordan-Wigner transformation:
\begin{equation}
\begin{split}
\hat a_j&=e^{-i\frac{\pi}{2}\sum^{2j-2}_{l=1}(\hat\sigma^z_l+1)}\hat\sigma^-_{2j-1},\\
\hat b_j&=e^{-i\frac{\pi}{2}\sum^{2j-1}_{l=1}(\hat\sigma^z_l+1)}\hat\sigma^-_{2j}.
\end{split}
\end{equation}
For example, the Rice-Mele Hamiltonian (\ref{HRM}) in the spin language becomes 
\begin{equation}
\begin{split}
\hat H_{\rm RM}=&-\sum_j(J_1\hat\sigma^+_{2j}\hat\sigma^-_{2j-1}+J_2\hat\sigma^+_{2j+1}\hat\sigma^-_{2j}+{\rm H.c.})\\
&-\sum_j\frac{\Delta}{2}(\hat\sigma^z_{2j-1}-\hat\sigma^z_{2j}),
\end{split}
\end{equation}
which describes a spin chain with anisotropic spin-flip coupling and subjected to a staggered magnetic field in the $z$ direction. Here we have used the identity $e^{-i\frac{\pi}{2}(\hat\sigma^z_j+1)}\hat\sigma^-_j=\hat\sigma^-_j$, since the state of the $j$th site must be $|\downarrow\rangle$ or vanish after the operation $\hat\sigma^-_j$. In the specific case $(J_1,J_2,\Delta)=(J_0,0,\Delta_0)$, the ground state is asymmetrically dimerized:
\begin{equation}
|\Psi\rangle=\bigotimes^L_{j=1}(\cos\frac{\theta_0}{2}|\uparrow_{2j-1}\downarrow_{2j}\rangle+\sin\frac{\theta_0}{2}|\downarrow_{2j-1}\uparrow_{2j}\rangle),
\label{Psi0}
\end{equation}
where $\theta_0\in(0,\pi)$ is determined from $\Delta_0=J_0\cot\theta_0$. Equation (\ref{Psi0}) can be rewritten into the following MPS form (see Fig.~\ref{figS6}(a)):
\begin{equation}
|\Psi\rangle=\sum_{\{s_j\}}{\rm Tr}[\Gamma^A_{s_1}\Lambda^A\Gamma^B_{s_2}\Lambda^B
...
\Gamma^B_{s_{2L}}\Lambda^B]|s_1s_2...
s_{2L}\rangle,
\label{MPS}
\end{equation}
where the sum runs over all the possible spin configurations $s_j=\uparrow,\downarrow$ ($j=1,2,...,2L$) and
\begin{equation}
\begin{split}
\Lambda^A&=\begin{bmatrix} \cos\frac{\theta_0}{2} & 0 \\ 0 & \sin\frac{\theta_0}{2}  \end{bmatrix},\;\Lambda^B=[1],\\
\Gamma^B_\uparrow=\begin{bmatrix} 0 \\ 1 \end{bmatrix},\;
\Gamma^B_\downarrow&=\begin{bmatrix} 1 \\ 0 \end{bmatrix},\;
\Gamma^A_\uparrow=\begin{bmatrix} 1 & 0 \end{bmatrix},\;
\Gamma^A_\downarrow=\begin{bmatrix} 0 & 1 \end{bmatrix}.
\end{split}
\label{iniGL}
\end{equation}

After the quench of the Hamiltonian, we can numerically calculate the MPS form of $|\Psi(t)\rangle=e^{-i\hat H't}|\Psi\rangle$ by using, e.g., the infinite time-evolving block decimation (iTEBD) algorithm \cite{Vidal2007} if we work in the thermodynamic  (infinite-$L$) limit. However, if the time-development operator is factorized as 
\begin{equation}
\hat U=\bigotimes_j\hat U_j, 
\label{Ufac}
\end{equation}
with $\hat U_j$ only acting on the spins at the $2j$th and the $(2j+1)$th sites (see Fig.~\ref{figS6}(b)), it is possible to analytically obtain the MPS form of $|\tilde\Psi\rangle=\hat U|\Psi\rangle$:
\begin{equation}
|\tilde\Psi\rangle=\sum_{\{s_j\}}{\rm Tr}[\tilde\Gamma^A_{s_1}\Lambda^A\tilde\Gamma^B_{s_2}\tilde\Lambda^B\tilde\Gamma^B_{s_{2L}}\tilde\Lambda^B]|s_1s_2...
s_{2L}\rangle,
\end{equation}
where the matrix ingredients $\tilde\Gamma^{A,B}$ and $\tilde\Lambda_B$ are related to those in Eq.~(\ref{MPS}) via \cite{Vidal2007}
\begin{equation}
\Lambda^A\tilde\Gamma^B_{s_1}\tilde\Lambda^B\tilde\Gamma^A_{s_2}\Lambda^A=U^{s_1s_2}_{s'_1s'_2}\Lambda^A\Gamma^B_{s'_1}\Lambda^B\Gamma^A_{s'_2}\Lambda^A,
\label{GLmap}
\end{equation}
where $U^{s_1s_2}_{s'_1s'_2}$ is the matrix element of a single block $\hat U_j$. 
If we are only interested in the inter-unit-cell ES, it suffices to find the singular values of 
the left-hand side of Eq.~(\ref{GLmap}), i.e.,
\begin{equation}
\begin{bmatrix} 
\Lambda^A\tilde\Gamma^B_\uparrow\tilde\Lambda^B\tilde\Gamma^A_\uparrow\Lambda^A & 
\Lambda^A\tilde\Gamma^B_\uparrow\tilde\Lambda^B\tilde\Gamma^A_\downarrow\Lambda^A \\ \Lambda^A\tilde\Gamma^B_\downarrow\tilde\Lambda^B\tilde\Gamma^A_\uparrow\Lambda^A & 
\Lambda^A\tilde\Gamma^B_\downarrow\tilde\Lambda^B\tilde\Gamma^A_\downarrow\Lambda^A 
\end{bmatrix}.
\label{SVDm}
\end{equation}

Equation (\ref{Ufac}) is satisfied for a general \emph{flat-band} quench in the Rice-Mele model:
\begin{equation}
(J_1,J_2,\Delta)=(J_0,0,\Delta_0)\to(0,J,\Delta),
\end{equation}
which implies 
\begin{equation}
\begin{split}
U^{\downarrow\downarrow}_{\downarrow\downarrow}&=U^{\uparrow\uparrow}_{\uparrow\uparrow}=\phi,\;\;\;\;\;\;\;\;\;\;\;\;\;\;\;\;
U^{\uparrow\downarrow}_{\downarrow\uparrow}=U^{\downarrow\uparrow}_{\uparrow\downarrow}=i\sin\theta\sin\phi,\\
U^{\uparrow\downarrow}_{\uparrow\downarrow}&=\cos\phi+i\cos\theta\sin\phi,\;\;U^{\downarrow\uparrow}_{\downarrow\uparrow}=\cos\phi-i\cos\theta\sin\phi,
\end{split}
\label{Udduu}
\end{equation}
where $\phi=\sqrt{J^2+\Delta^2}t$ and $\theta\in(0,\pi)$ is defined from $\Delta=J\cot\theta$. Combining Eqs.~(\ref{Udduu}) and (\ref{iniGL}), we can explicitly compute the matrix elements in Eq.~(\ref{SVDm}) to obtain
\begin{widetext}
\begin{equation}
\begin{bmatrix} 0 & 0 & i\cos^2\frac{\theta_0}{2}\sin\theta\sin\phi & 0 \\ \sin\frac{\theta_0}{2}\cos\frac{\theta_0}{2} & 0 & 0 & \sin^2\frac{\theta_0}{2}(\cos\phi+i\cos\theta\sin\phi) \\ \cos^2\frac{\theta_0}{2}(\cos\phi-i\cos\theta\sin\phi) & 0 & 0 & \sin\frac{\theta_0}{2}\cos\frac{\theta_0}{2} \\ 0 & i\sin^2\frac{\theta_0}{2}\sin\theta\sin\phi & 0 & 0 \end{bmatrix},
\label{gachiSVDm}
\end{equation}
\end{widetext}
The four singular values of the above matrix (\ref{gachiSVDm}) are found to be
\begin{equation}
\begin{split}
\frac{1}{2}\left(\sqrt{1-\sin^2\theta\cos^2\theta_0\sin^2\phi}\pm\sqrt{1-\sin^2\theta\sin^2\phi}\right),\\
\frac{1}{2}\sin\theta(1\pm\cos\theta_0)|\sin\phi|,\;\;\;\;\;\;\;\;\;\;\;\;\;\;\;\;\;\;\;\;\;\;
\end{split}
\label{LamB}
\end{equation}
which constitute the eigenvalues of $\tilde\Lambda_B$. The single-particle ES can subsequently be determined from Eq.~(\ref{LamB}) via the relation between many-body ES and single-particle ES given in Eq.~(\ref{mbl}) in the main text (see also Ref.~\cite{Fidkowski2010}). The result turns out to be
\begin{equation}
\begin{split}
&\xi_n=\frac{1}{2}\left[1\pm\cos\theta_0\sin^2\theta\sin^2\phi\right.\\
&\left.\pm\sqrt{(1-\sin^2\theta\sin^2\phi)(1-\cos^2\theta_0\sin^2\theta\sin^2\phi)}\right].
\end{split}
\label{anaxi}
\end{equation}

Finally, let us discuss a specific case in which $\theta=\theta_0$, namely the quench shown in Fig.~\ref{figS5}(a). In this case, the single-particle-ES gap reaches its minimum at $\phi=\frac{\pi}{2}$, where
\begin{equation}
\xi_n=\frac{1}{2}\left[1\pm\cos\theta\left(\sin^2\theta\pm\sqrt{1-\cos^2\theta\sin^2\theta}\right)\right],
\end{equation}
implying a tiny gap 
\begin{equation}
\delta\xi=\frac{1}{2}\cos^3\theta(1+\cos^2\theta)\simeq\frac{\Delta^3}{2J^3}
\end{equation}
for a small $\Delta$. This is consistent with the small gap found in the inset of Fig.~\ref{figS5}(b). If we further set $\theta=\theta_0=\frac{\pi}{2}$, Eq.~(\ref{anaxi}) becomes
\begin{equation}
\xi_n=\frac{1}{2}[1\pm\cos(Jt)],
\end{equation}
which reproduces the sinusoidal oscillation observed in Fig.~\ref{fig1}(c) (red curve) 
in the main text.

\subsection{iTEBD for translation-invariant interacting systems}
As mentioned in the previous subsection and at the end of the main text, the iTEBD algorithm is a general numerical method for calculating the quantum dynamics in translation-invariant one-dimensional interacting systems in the thermodynamic limit. Since iTEBD is based on the MPS representation, we can directly read out the many-body ES from $\Lambda$ matrices. We should emphasize that the many-body ES $\lambda_\alpha$ obtained from iTEBD corresponds to the \emph{open-boundary condition}. The ES for the periodic-boundary condition can thus be obtained as $\lambda_\alpha\lambda_\beta$ for all $\alpha$ and $\beta$. We should also mention that the iTEBD is reliable only for short-time dynamics due to the \emph{exponential} growth of the bond dimension (effective size of $\Lambda^{A,B}$ or $\Gamma^{A,B}$) which is implied by the linear growth of entanglement entropy.

As an example, we consider an interacting SSH chain
\begin{equation}
\begin{split}
&\hat H_{\rm ISSH}=-\sum_j(J_1\hat b^\dag_j\hat a_j+J_2\hat a^\dag_{j+1}\hat b_j+{\rm H.c.})\\
&-\sum_j\left(\hat n^b_j-\frac{1}{2}\right)\left[V_1\left(\hat n^a_j-\frac{1}{2}\right)+V_2\left(\hat n^a_{j+1}-\frac{1}{2}\right)\right],
\end{split}
\label{HISSH}
\end{equation}
where $\hat n^b_j=\hat b^\dag_j\hat b_j$ and $\hat n^a_j=\hat a^\dag_j\hat a_j$ are the particle-number operators. After the Jordan-Wigner transformation, Eq.~(\ref{HISSH}) can be rewritten into the following Hamiltonian that describes a staggered XXZ chain:
\begin{equation}
\begin{split}
\hat H_{\rm XXZ}=&-\sum_j(J_1\hat\sigma^+_{2j}\hat\sigma^-_{2j-1}+J_2\hat\sigma^+_{2j+1}\hat\sigma^-_{2j}+{\rm H.c.})\\
&-\sum_j(J^z_1\hat\sigma^z_{2j}\hat\sigma^z_{2j-1}+J^z_2\hat\sigma^z_{2j+1}\hat\sigma^z_{2j}),
\end{split}
\label{HXXZ}
\end{equation}
where $J^z_{1,2}=V_{1,2}/4$ and they are chosen to satisfy $J^z_1=J^z_2=J_z$ in the following calculations. The block time-evolution operator thus reads
\begin{equation}
\begin{split}
[U_{1,2}(t)]^{\downarrow\downarrow}_{\downarrow\downarrow}&=[U_{1,2}(t)]^{\uparrow\uparrow}_{\uparrow\uparrow}=e^{iJ_zt},\\
[U_{1,2}(t)]^{\uparrow\downarrow}_{\downarrow\uparrow}&=[U_{1,2}(t)]^{\downarrow\uparrow}_{\uparrow\downarrow}=ie^{-iJ_zt}\sin(J_{1,2}t),\\
[U_{1,2}(t)]^{\uparrow\downarrow}_{\uparrow\downarrow}&=[U_{1,2}(t)]^{\downarrow\uparrow}_{\downarrow\uparrow}=e^{-iJ_zt}\cos(J_{1,2}t),
\end{split}
\label{UXXZ}
\end{equation}
from which the updated MPS form after a short time period $\delta t$ is determined as (see Fig.~\ref{figS6}(c))
\begin{equation}
\begin{split}
\Lambda^B\Gamma^{A\star}_{s_1}\tilde\Lambda^A\Gamma^{B\star}_{s_2}\Lambda^B=[U_1(\delta t)]^{s_1s_2}_{s'_1s'_2}\Lambda^B\Gamma^A_{s'_1}\Lambda^A\Gamma^B_{s'_2}\Lambda^B,\\
\tilde\Lambda^A\tilde\Gamma^B_{s_1}\tilde\Lambda^B\tilde\Gamma^A_{s_2}\tilde\Lambda^A=[U_2(\delta t)]^{s_1s_2}_{s'_1s'_2}\tilde\Lambda^A\Gamma^{B\star}_{s'_1}\Lambda^B\Gamma^{A\star}_{s'_2}\tilde\Lambda^A.
\end{split}
\end{equation}
In practical calculations, to suppress the numerical errors stemming from a finite $\delta t$, we reorder the block time-evolution operators to construct the 4th-order sequence \cite{Stewart1999}
\begin{equation}
\begin{split}
\hat U^{(4)}&=\hat U_1\hat U^2_2\hat U^2_1\hat U^\dag_2\hat U^\dag_1(\hat U_2\hat U_1)^3\hat U^2_2\hat U_1\\
&\times\hat U_1\hat U^2_2(\hat U_1\hat U_2)^3\hat U^\dag_1\hat U^\dag_2\hat U^2_1\hat U^2_2\hat U_1
\end{split}
\end{equation}
such that $\hat U^{(4)}=e^{-12i\delta t \hat H_{\rm XXZ}+o(\delta t^4)}$. The ground state of the initial Hamiltonian can be obtained by the imaginary time development based on the same iTEBD algorithm.

\begin{figure}
\begin{center}
        \includegraphics[width=8.5cm, clip]{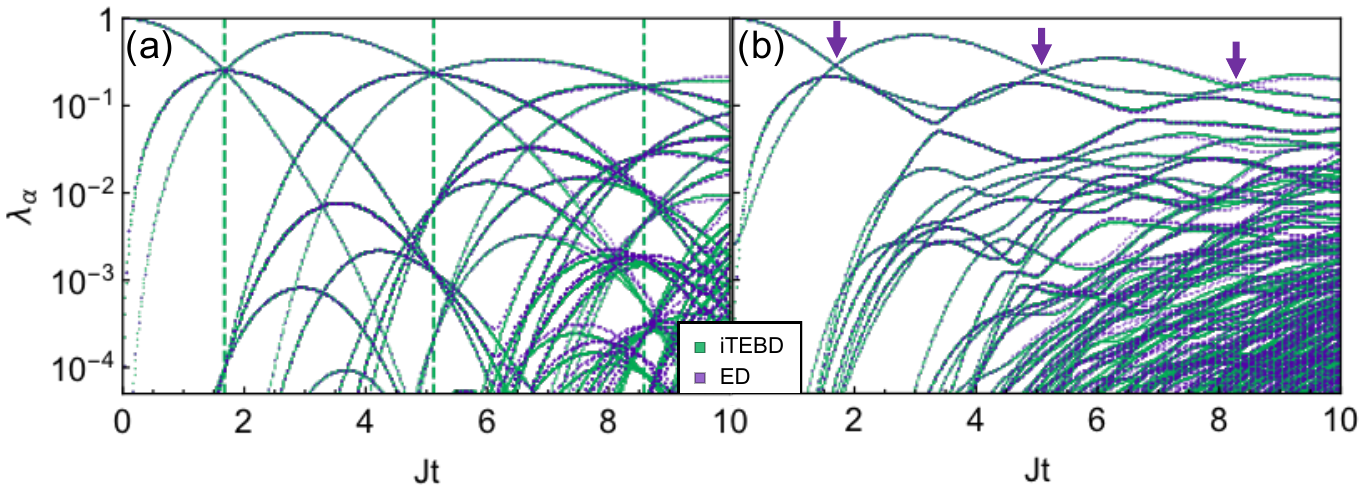}
      \end{center}
   \caption{Dynamics of the half-chain many-body ES in the staggered XXZ model (\ref{HXXZ}) after a quench $(J_1,J_2)=(J,0)\to(0.4J,J)$. (a) Without interaction ($J_z=0$), global ES degeneracies emerge at certain times (indicated by dashed lines). The bond dimension in iTEBD calculation is $\chi=250$, and the system size in the exact-diagonalization (ED) calculation is  $L=8$. (b) With interaction ($J_z=0.2J$), degeneracies survive in the largest two $\lambda_\alpha$'s at certain times (indicated by arrows). Here $\chi=500$ and $L=8$.}
   \label{figS7}
\end{figure}

Figure \ref{figS7} shows the many-body-ES dynamics obtained by iTEBD for both noninteracting and interacting cases. Similar to the transverse-field Ising model, in the noninteracting case (Fig.~\ref{figS7}(a)), we find global ES degeneracies at certain times, which 
exactly coincide with those for single-particle ES crossings at $\xi=\frac{1}{2}$ in Fig.~\ref{fig1}(c) (the curve for $J'=0.4J$) in the main text. In the interacting case (Fig.~\ref{figS7}(b)), although the global degeneracies disappear, the crossings of the largest two ES values (instantaneous ground-state degeneracy of the many-body entanglement Hamiltonian) survive. This observation implies the robustness of ES crossings against interaction.

\begin{figure}
\begin{center}
        \includegraphics[width=8.5cm, clip]{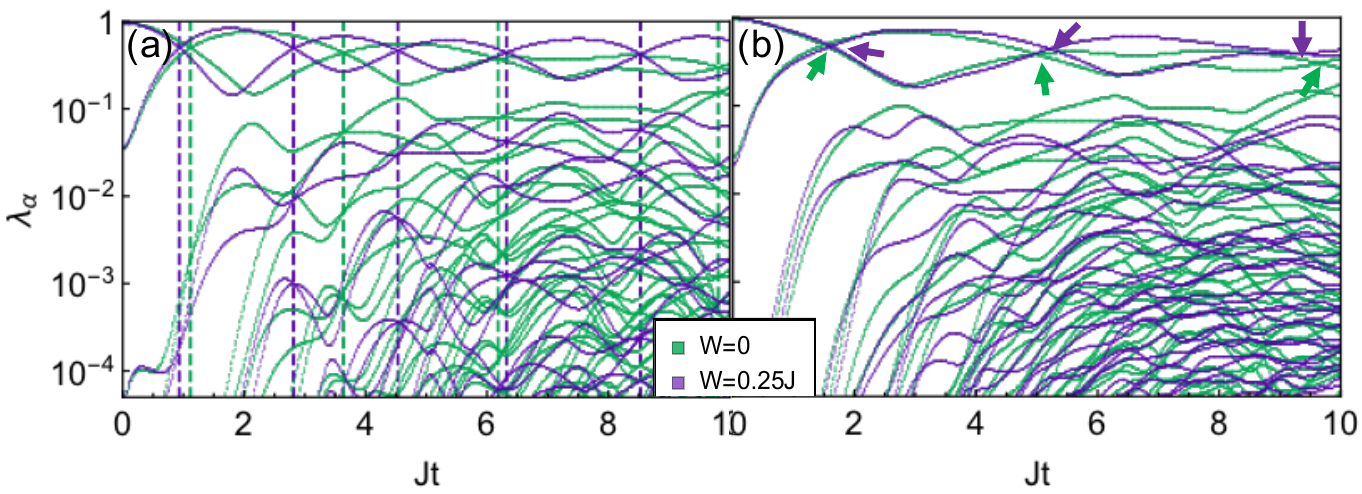}
      \end{center}
   \caption{Dynamics of the half-chain many-body ES in the generalized Ising model (\ref{GIsing}) after a quench $\bar B_j=1.5J\to0.5J$. The magnetic field $B_j$ is randomly sampled from a uniform distribution over $[\bar B_j-W,\bar B_j+W]$ with $W=0.25J$, and the system size is $L=12$. (a) For a nearest-neighbor interaction $J_{jl}=J_j\delta_{l,j+1}$, where $J_j$ obeys the uniform distribution over $[J-0.6W,J+0.6W]$, global ES degeneracies emerge at certain times (indicated by dashed lines). (b) For a long-range interaction $J_{jl}=J/|j-l|^\alpha$ with $\alpha=3$, degeneracies survive in the largest two $\lambda_\alpha$'s at certain times (indicated by arrows).}
   \label{figS8}
\end{figure}

\subsection{Exact diagonalization for arbitrary small-size systems}
\label{EDAS}
If we are interested in the many-body-ES dynamics in a relatively small system, we can always use numerical exact diagonalization even if the system is disordered or/and interacting. 

In Fig.~\ref{figS7}, we also present the results for a small XXZ chain with $2L=16$ spins, which turn out to be a rather good approximation of the iTEBD results in the thermodynamic limit. Note that the global degeneracies in the many-body ES persist (although slightly shifted compared to those in the thermodynamic limit) even in a finite-size system, as has already been observed in Fig.~\ref{fig3}(b) in the main text. This is understandable since the crossings originate from the topology on the momentum-time space, of which the time degree of freedom is always continuous.

\begin{figure*}
\begin{center}
        \includegraphics[width=17.6cm, clip]{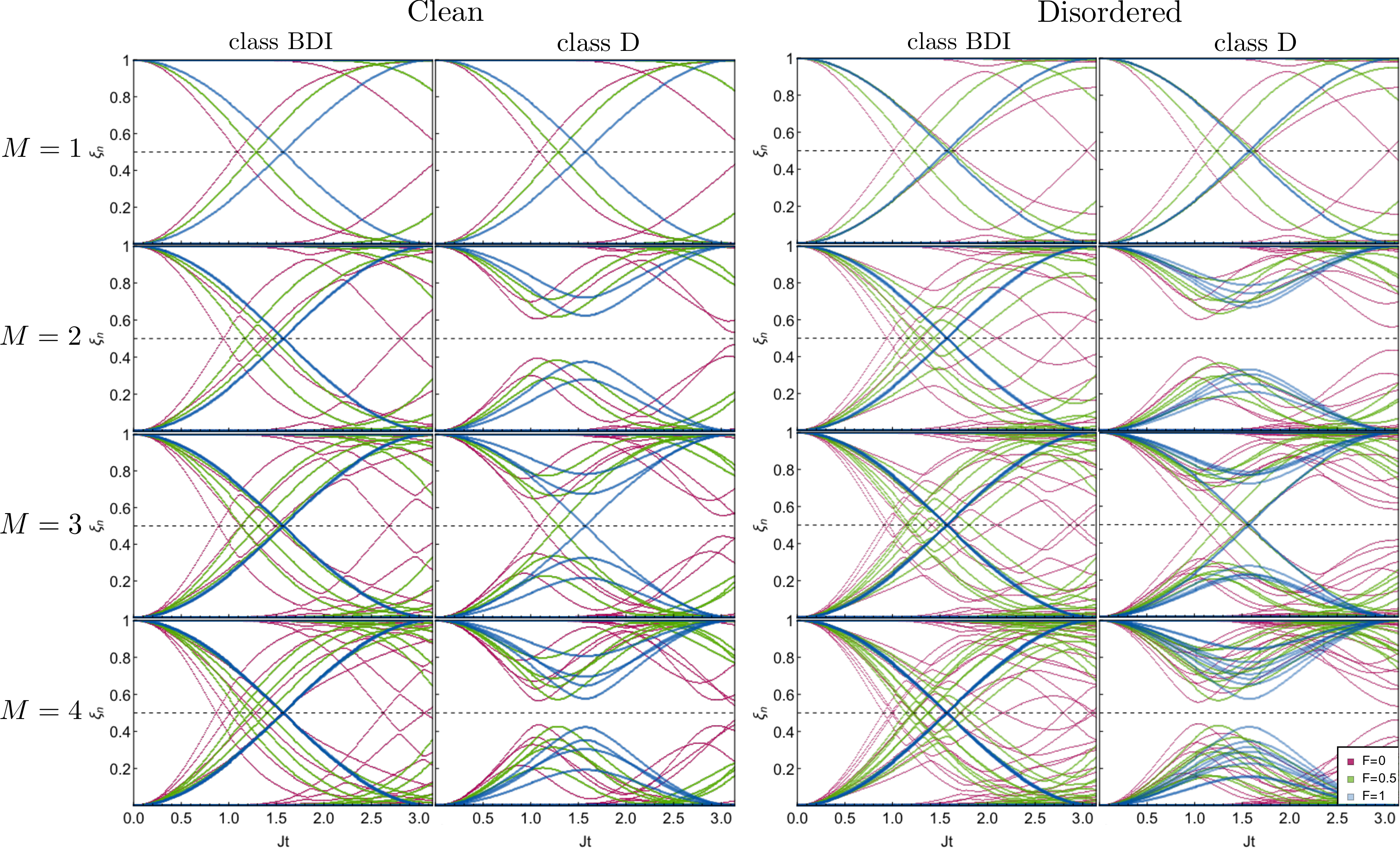}
      \end{center}
   \caption{Single-particle ES dynamics in $M$ copies of SSH chains with interchain couplings (see Fig.~\ref{fig3}(a) and (b) in the main text) that either respect both TRS and PHS (class BDI, see Eq.~(\ref{HBDI})) or break the TRS alone (class D, see Eq.~(\ref{HD})). The quench parameters are chosen to be the same as those in the main text, i.e., $(\bar J_1,\bar J_2,\bar J_c)=(0,1.5J,0)\to(1.5J,0.5J,0.5J)$ for class BDI and $(\bar J_1,\bar J_2,\bar J_c)=(0,1.5J,0)\to(1.5J,0.5J,J)$ for class D. The disorder realization of $J_1$ and $J_2$ is set to be the same for class BDI and class D, so their ES dynamics agree for $M=1$ in the disordered case. The length of a single chain is chosen to be $L=\frac{120}{M}$.}
   \label{figS9}
\end{figure*}

We also perform further numerical calculations on the ES dynamics in the generalized Ising model: \begin{equation}
\hat H=\sum_{j<l}J_{jl}\hat\sigma^x_j\hat\sigma^x_l+\sum_jB_j\hat\sigma^z_j.
\label{GIsing}
\end{equation}
Figure.~\ref{figS8}(a) shows the results for a nearest-neighbor interaction $J_{jl}=J_j\delta_{l,j+1}$, so the system can be mapped into a disordered Kitaev chain belonging to class D \cite{Kitaev2001}. 
It turns out that the global degeneracy survives (although their positions change dramatically) the disorder in both $J_j$ and $B_j$. Figure.~\ref{figS8}(a) shows the results for a long-range power-law interaction $J_{jl}=J/|j-l|^\alpha$, which is relevant to trapped ions \cite{Zhang2017b} and Rydberg-atom arrays \cite{Bernien2017}. Due to the fact that the system can no longer be mapped onto free fermions, similarly to Fig.~\ref{figS7}(b), global degeneracies disappear, but the crossing between the largest two ES values survive. Such a crossing is shown to be robust against disorder in $B_j$.

\section{Further numerical evidence on the $\mathbb{Z}_2$ topological index}
In this section, we perform extensive numerical calculations of the ES dynamics with $\mathbb{Z}_2$ characterization, including class D and class DIII.

\subsection{Class D}

Based on the quench in a single SSH chain, which realizes the generator $1$ in the K-group (see Table~\ref{tableS1}), we can obtain any element $M\in\mathbb{Z}^+$ by simply quenching $M$ copies of SSH chains. Correspondingly, the number of ES crossings at $\frac{1}{2}$ is multiplied by $M$. To demonstrate the topological nature of these ES crossings, we can randomly deform each chain by introducing randomness in hopping amplitudes:
\begin{equation}
\hat H_\alpha=-\sum_j(J_{1,j\alpha}\hat b^\dag_{j\alpha}\hat a_{j\alpha}+J_{2,j\alpha}\hat a^\dag_{j+1,\alpha}\hat b_{j\alpha}+{\rm H.c.}),
\end{equation}
and further randomly couple these SSH chains:
\begin{equation}
\hat H_{\rm BDI}=\sum^M_{\alpha=1}\hat H_\alpha-\sum^{M-1}_{\alpha=1}\sum_j(J_{c,j\alpha}\hat b^\dag_{j,\alpha+1}\hat a_{j\alpha}+{\rm H.c.}).
\label{HBDI}
\end{equation}
Here $\hat a_{j\alpha}$ ($\hat b_{j\alpha}$) is the annihilation operator of a particle at sublattice A in the $j$th unit cell of the $\alpha$th SSH chain. Note that both TRS and PHS are preserved in Eq.~(\ref{HBDI}), so that the entire system still belongs to class BDI. If we choose a different form of interchain hopping as described by the Hamiltonian
\begin{equation}
\hat H_{\rm D}=\sum^M_{\alpha=1}\hat H_\alpha+\sum^{M-1}_{\alpha=1}\sum_j(iJ_{c,j\alpha}\hat a^\dag_{j,\alpha+1}\hat a_{j\alpha}+{\rm H.c.}),
\label{HD}
\end{equation}
we can break the TRS alone so that the symmetry class of the entire system reduces to class D.

In the main text we have presented the ES dynamics for $M=3$ coupled SSH chains, with (\ref{HBDI}) or without (\ref{HD}) TRS breaking. 
After band flattening, we find $2M=6$ ES crossings in a period in the former case, while there are only $2$ crossings in the latter case. Such an observation strongly suggests a reduction of topological number $\mathbb{Z}\to\mathbb{Z}_2$. 

To further support such a dynamical $\mathbb{Z}_2$ reduction, we perform extensive numerical calculations for $M=1,2,3,4$ coupled SSH rings in both class BDI and class D, with or without disorder. As shown in Fig.~\ref{figS9}, we clearly find $2M$ and $2(M\;{\rm mod}\;2)$ ES crossings in a single period in the flat band limit, no matter whether the system is clean or disordered. These crossings persist when the bandwidth becomes finite and the periodicity disappears, at least for the parameter choices and a short time interval as shown in Fig.~\ref{figS9}. 

On the other hand, for larger interchain coupling, such as $J_c=1$ for $M=3$ chains in class BDI, we find a pair annihilation of ES crossings even in the clean limit when we gradually change $F$ from $1$ to $0$. This observation implies that \emph{not} all the ES crossings in the flat-band limit can survive when the bandwidth becomes finite. Indeed, it is already known in equilibrium systems that too strong disorder can kill the edge modes and drive a topological-to-trivial transition \cite{Prodan2016}. This is why we emphasize ``provided that the disorder in frequency domain due to band nonflatness is not so strong" when discussing the multiplication of ES crossings in the main text.

\subsection{Class DIII}
We numerically demonstrate that the $\mathbb{Z}_2$ topological index for class DIII also has an intuitive implication on the ES dynamics. In fact, by preparing two copies (with opposite spins) of the class D SSH chains, we have already obtained a model in class DIII, which certainly exhibits the $\mathbb{Z}_2$ reduction with doubled degeneracies in the ES crossings (due to the Kramers degeneracy). Note that once TRS alone is broken, the system reduces to class D and is always trivial.

To make the demonstration nontrivial, we can add a spin-orbit coupling term that respects both TRS and PHS:
\begin{equation}
\begin{split}
\hat H_\alpha=&-\sum_{j,s}(J_{1,j\alpha}\hat b^\dag_{j\alpha s}\hat a_{j\alpha s}+J_{2,j\alpha}\hat a^\dag_{j+1,\alpha s}\hat b_{j\alpha s}+{\rm H.c.})\\
&+\sum_{j}[J_{{\rm c},j\alpha}(\hat b^\dag_{j\alpha\downarrow}\hat a_{j\alpha\uparrow}-\hat a^\dag_{j\alpha\downarrow}\hat b_{j\alpha\uparrow})+{\rm H.c.}],
\end{split}
\label{sgDIII}
\end{equation}
where $s=\uparrow,\downarrow$ denotes the spin degree of freedom. Denoting $\hat{\boldsymbol{c}}_j=[\hat a_{j\uparrow},\hat b_{j\uparrow},\hat a_{j\downarrow},\hat b_{j\downarrow}]^{\rm T}$, for an arbitrary $z\in\mathbb{C}$, the TRS and the PHS act like
\begin{equation}
\begin{split}
\mathcal{T}(z\hat{\boldsymbol{c}}_j)\mathcal{T}^{-1}&=z^*(i\sigma^y\otimes\sigma^0)\hat{\boldsymbol{c}}_j,\\
\mathcal{C}(z\hat{\boldsymbol{c}}_j)\mathcal{C}^{-1}&=z^*(\sigma^0\otimes\sigma^z)\hat{\boldsymbol{c}}_j.
\end{split}
\end{equation}
We can couple $M$ copies of such class DIII chains as
\begin{equation}
\begin{split}
\hat H=\sum^M_{\alpha=1}\hat H_\alpha&+\sum^{M-1}_{\alpha=1}\sum_{j}[(iJ_{{\rm cc},j\alpha}\hat a^\dag_{j,\alpha+1,\uparrow}\hat a_{j\alpha\uparrow}\\
&-iJ_{{\rm cc},j\alpha}\hat a^\dag_{j,\alpha+1,\downarrow}\hat a_{j\alpha\downarrow})+{\rm H.c.}],
\end{split}
\label{MDIII}
\end{equation}
which respects both TRS and PHS.

\begin{figure}
\begin{center}
        \includegraphics[width=8cm, clip]{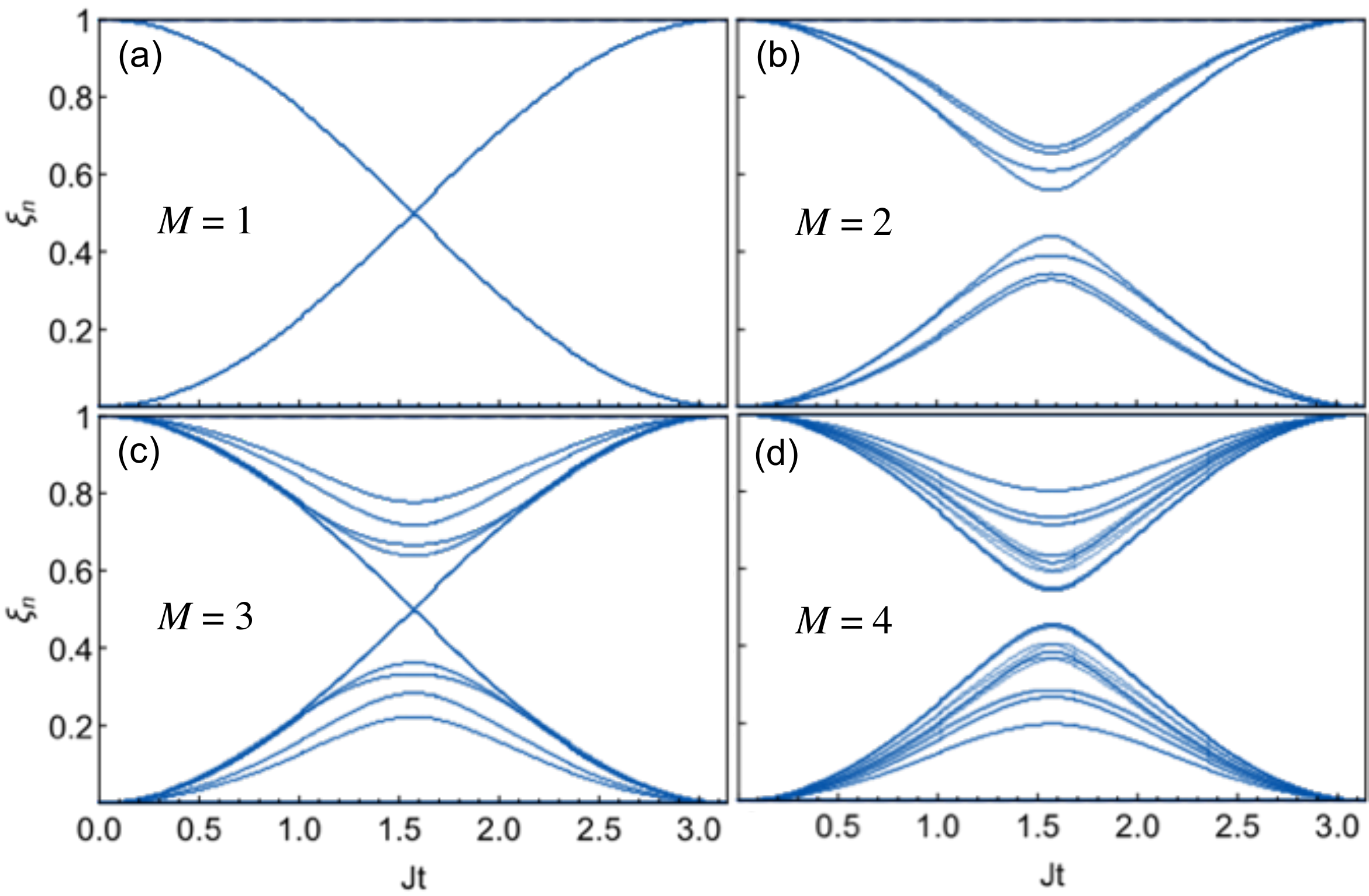}
      \end{center}
   \caption{Entanglement-spectrum dynamics in a disordered class DIII system (\ref{MDIII}) after band flattening. (a)-(d) correspond to the case of $M=1\sim 4$ coupled four-band class DIII chains (\ref{sgDIII}). Note that a four-fold-degenerate (no) crossing survives for an odd (even) M, implying a robust $\mathbb{Z}_2$ topological characterization. The length of a single chain is chosen to be $L=\frac{96}{M}$.}
   \label{figS10}
\end{figure}

We calculate the ES dynamics after a quench in the model given in Eq.~(\ref{MDIII}). All the parameters are randomly sampled from $J_\mu\in[0.6\bar J_\mu,1.4\bar J_\mu]$, and the quench protocol is chosen to be
\begin{equation}
(\bar J_1,\bar J_2,\bar J_{\rm c},\bar J_{\rm cc})=(1.5J,0,0,0)
\to(0,1.5J,0.4J,0.8J).
\end{equation}
The numerical results are shown in Fig.~\ref{figS10} for $M=1\sim4$ coupled chains. Remarkably, we find that there is a four-degenerate (this comes from the periodic-boundary condition and the Kramers degeneracy) crossing for $M=1,3$ while no crossing for $M=2,4$. Such an observation is consistent with the $\mathbb{Z}_2$ classification predicted by the K-theory (see Table~\ref{tableS1}).

\section{Comment on the relationship between entanglement-spectrum crossings and dynamical phase transitions}
While an interesting conjecture that the ES crossings might be related to the dynamical phase transitions \cite{Heyl2013} is made in Ref.~\cite{Chiara2014}, here we show that these two concepts are not equivalent because of the following two aspects: (i) There can be a dynamical phase transition without any ES crossing; (ii) even if both occur, the time at which an ES crossing occurs may not coincide with that at which a singularity in the dynamical free-energy density emerges.

Let us begin by reviewing how to calculate the dynamical free-energy density \cite{Heyl2013}
\begin{equation}
f(t)=-\lim_{L\to\infty}\frac{1}{L}\ln|\langle\Psi|e^{-i\hat H't}|\Psi\rangle|^2
\label{DFE}
\end{equation}
in free-fermion lattice systems. Following Eq.~(\ref{Psifac}), we can also factorize $|\Psi(t)\rangle=e^{-i\hat H't}|\Psi\rangle$ as 
\begin{equation}
|\Psi(t)\rangle=\prod_k\boldsymbol{c}^\dag_k\boldsymbol{u}(k,t)|{\rm vac}\rangle.
\end{equation}
In terms of the Bloch vector $\boldsymbol{u}(k,t)$ ($\boldsymbol{u}(k,0)=\boldsymbol{u}(k)$), Eq.~(\ref{DFE}) can be rewritten as
\begin{equation}
f(t)=-\int^\pi_{-\pi}\frac{dk}{2\pi}\ln|(\boldsymbol{u}(k))^\dag\boldsymbol{u}(k,t)|^2,
\end{equation}
where $\lim_{L\to\infty}\frac{1}{L}\sum_k=\int^\pi_{-\pi}\frac{dk}{2\pi}$ has been used. For 
two-band systems, the fidelity $|(\boldsymbol{u}(k))^\dag\boldsymbol{u}(k,t)|^2$ can easily be evaluated as
\begin{equation}
{\rm Tr}\left[\frac{1+\boldsymbol{n}(k,t)\cdot\boldsymbol{\sigma}}{2}\frac{1+\boldsymbol{n}(k)\cdot\boldsymbol{\sigma}}{2}\right]=\frac{1}{2}[1+\boldsymbol{n}(k,t)\cdot\boldsymbol{n}(k)],
\end{equation}
where $\boldsymbol{n}(k,t)\equiv-\frac{\boldsymbol{d}(k,t)}{d(k)}$. Using Eq.~(\ref{genfor}), we  obtain the general formula \cite{Vajna2015}
\begin{equation}
\begin{split}
f(t)=&-\int^\pi_{-\pi}\frac{dk}{2\pi}\ln[\cos^2(d'(k)t)\\
&+(\boldsymbol{n}(k)\cdot\boldsymbol{n}'(k))^2\sin^2(d'(k)t)].
\end{split}
\label{2bddfe}
\end{equation}
It can be inferred from Eq.~(\ref{2bddfe}) that the singularity of $f(t)$ is contributed from those $k^*$ at which $\boldsymbol{n}(k^*)\cdot\boldsymbol{n}'(k^*)=0$ and emerges at critical times $t^*_n=(n-\frac{1}{2})\frac{\pi}{d'(k^*)}$ with $n\in\mathbb{Z}^+$ \cite{Heyl2013}.

\begin{figure}
\begin{center}
        \includegraphics[width=8.5cm, clip]{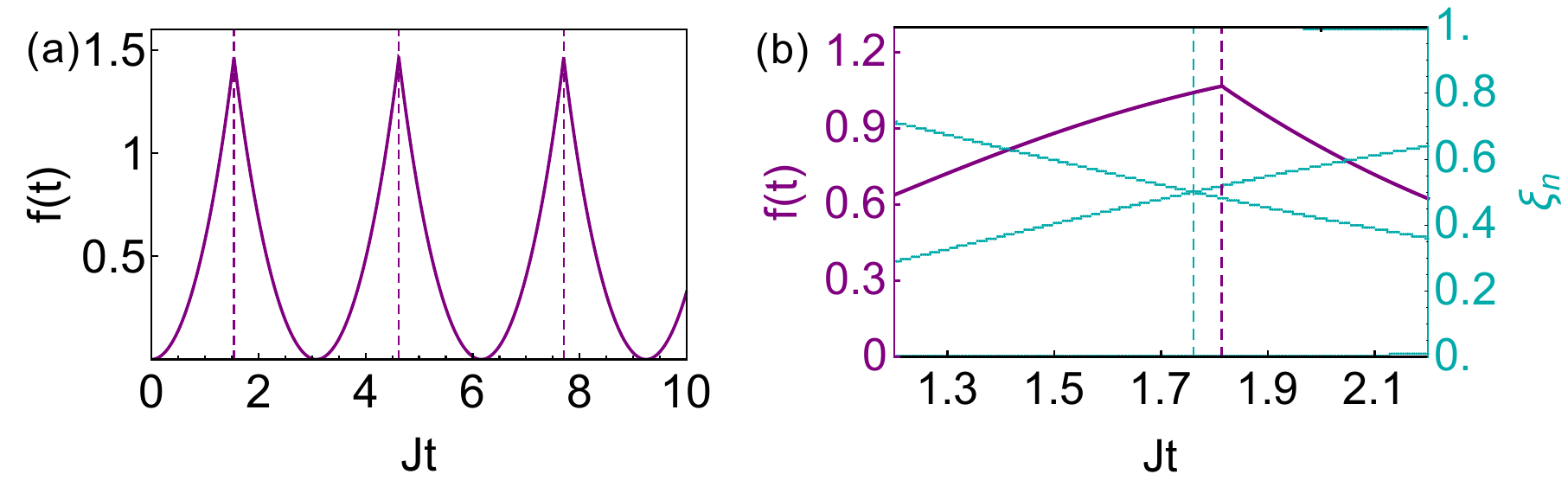}
      \end{center}
   \caption{(a) Dynamical free-energy density $f(t)$ (\ref{DFE}) for the quench in the Rice-Mele model with $J=0.2\Delta$. 
   The critical times (\ref{crt}) are indicated by dashed vertical lines. (b) ES dynamics (green) and $f(t)$ (purple) for the quench in the SSH model with $J'=0.5J$. 
   The first critical time (\ref{crt2}) does not exactly coincide with the first ES-crossing time. Note that there are always some ES values very close to $0$ or $1$.}
   \label{figS11}
\end{figure}

To demonstrate statement (i), 
it suffices to consider the quench in the Rice-Mele model illustrated in Fig.~\ref{figS5}(a), which does not exhibit ES crossings. However, it is easy to check that $\boldsymbol{n}(k)\cdot\boldsymbol{n}'(k)=0$ does have solutions $\cos k^*=-\frac{\Delta^2}{J^2}$, implying the emergence of cusps in $f(t)$ at the critical times
\begin{equation}
t^*_n=\frac{(n-\frac{1}{2})\pi}{\sqrt{J^2+\Delta^2}},\;\;n=1,2,\cdots.
\label{crt}
\end{equation}
As shown in Fig.~\ref{figS11}(a), we indeed find the singularities in the dynamical free-energy density.  

To demonstrate statement (ii), we can consider a quench in the SSH model (\ref{HSSH}) with $J'=0.5J$. In this case, $\boldsymbol{n}(k)\cdot\boldsymbol{n}'(k)=0$ has solutions $\cos k^*=-\frac{J'}{J}$ and the critical times are given by
\begin{equation}
t^*_n=\frac{(n-\frac{1}{2})\pi}{\sqrt{J^2-J'^2}},\;\;n=1,2,\cdots.
\label{crt2}
\end{equation}
As shown in Fig.~\ref{figS11}(b), we do find a cusp in $f(t)$ at the first critical time $t^*_1\simeq1.81J^{-1}$. On the other hand, by numerically calculating the single-particle ES dynamics, we find that the first ES crossing occurs at $t_1\simeq1.76J^{-1}$, which is close to but not equal to $t^*_1$.

\section{Experimental situations}

We briefly discuss how to experimentally measure the ES dynamics in the SSH model simulated by ultracold atoms and that in the Ising model simulated by trapped ions. It is worth mentioning that an interferometric scheme to directly measure the ES in an interacting ultracold atomic system has been proposed in Ref.~\cite{Pichler2016}, but not yet realized experimentally. Such a scheme is a generalization of Ref.~\cite{Daley2012}, which describes a method for measuring the R\'enyi entropy and has recently been realized in an optical lattice \cite{Greiner2015}.   

\subsection{Ultracold atoms}
Ultracold atoms in optical lattices provide an ideal platform to explore nonequilibrium quantum dynamics, which has been visualized in both real and momentum spaces \cite{Schmiedmayer2015}. Indeed, sudden quench and Bloch-state tomography have been achieved in the Haldane model implemented by ${}^{40}$K atoms \cite{Weitenberg2017,Weitenberg2018}. Here we apply these ideas and techniques to the SSH model.

\begin{figure}
\begin{center}
        \includegraphics[width=7.3cm, clip]{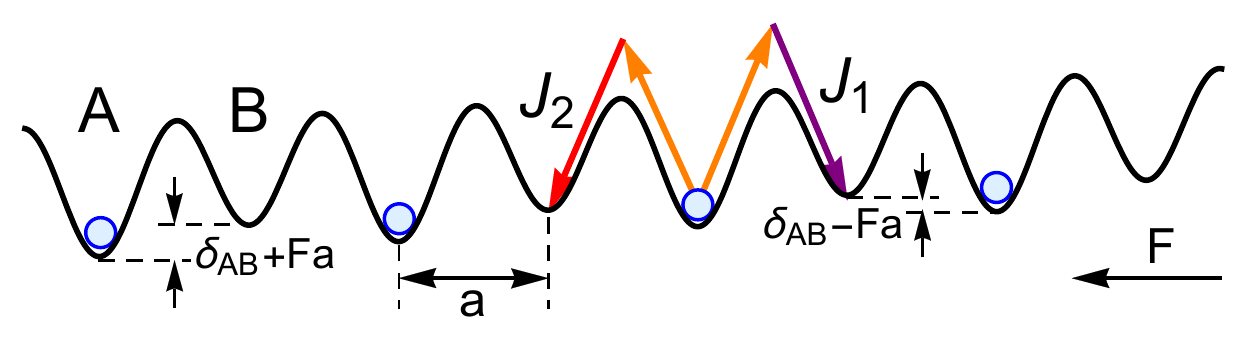}
      \end{center}
   \caption{(color online) Implementation of the SSH model based on laser-assistant tunneling with ultracold atoms in a tilted optical superlattice. The intra- and inter-unit-cell hoppings $J_1$ and $J_2$ can independently be controlled by using two pairs of Raman lasers with detunings $\delta_{\rm AB}\pm Fa$.}
   \label{figS12}
\end{figure}

While the SSH model and the corresponding Rice-Mele model have been realized in Refs.~\cite{Monika2013b,Monika2016b,Takahashi2016}, they are not suitable for studying quench dynamics since the Wannier functions change considerably after deforming the optical lattice, leading to unwanted excitations in higher bands. Also, an energy difference between A, B sublattices is needed to perform tomography \cite{Hauke2014}. Therefore, we use a superlattice with large energy offset $\delta_{\rm AB}$ between nearest neighbors (separated by $a$) and subjected to a uniform potential gradient $Fa\sum_j[(2j-1)\hat a^\dag_j\hat a_j+2j\hat b^\dag_j\hat b_j]$ (see Fig.~\ref{figS12}). Thanks to the potential gradient, it is possible to independently control the hopping parameters $J_1$ and $J_2$ by two pairs of Raman lasers with detunings $\delta_{\rm AB}\pm Fa$. For example, to realize the quench $(J_0,0)\to(0,J)$, we can suddenly switch off one laser assistant tunneling $J_1=J_0$ and switch on the other tunneling $J_2=J$. Note that it is easy to generalize to the Rice-Mele model by choosing imperfect Raman-laser detunings $\delta_{\rm AB}\pm Fa\mp\Delta$.

As for Bloch-state tomography, we follow the method in Ref.~\cite{Weitenberg2016} to suddenly switch off the potential gradient and Raman lasers and then perform time-of-flight measurements after waiting for a varying time up to a few times of $2\pi\delta^{-1}_{\rm AB}$. Unlike the case in Ref.~\cite{Weitenberg2016}, the potential gradient continuously shifts the quasimomenta during the quench dynamics, so the measured Bloch state at $k$ should be replaced by $k+Ft_{\rm q}$, with $t_{\rm q}$ being the time duration of the quench dynamics. In practice, we can apply an opposite potential gradient during time $t_{\rm q}$ to compensate for this effect. If the gradient comes from inhomogeneous Zeeman splitting, this can be done by globally flipping the atomic spin \cite{Monika2013b}. With the full information of Bloch states in hand, we can calculate the half-chain ES using Eq.~(\ref{Cjalb}), with $L$ depending on the number of data. 


\subsection{Trapped ions}
As mentioned in Sec.~\ref{EDAS},  the generalized Ising model (\ref{GIsing}) with a power-law long-range coupling can naturally be realized in a linear chain of trapped ions. With the ability to address individual ions, we can in principle engineer an arbitrary configuration of Ising coupling $J_{jl}$ by fine-tuning the parameters (frequency, phase and intensity) of Raman lasers \cite{Monroe2012}. A spatially inhomogeneous magnetic field can effectively be generated by site-dependent Stark shifts induced by additional lasers \cite{Monroe2016}.

To probe the ES dynamics after a sudden change of parameters, we can make use of the MPS tomography \cite{Cramer2010}, which has recently been achieved for up to $14$ ${}^{40}$Ca${}^+$ ions \cite{Blatt2017}. 
Indeed, the entanglement-entropy growth during the quench dynamics has been measured in Ref.~\cite{Blatt2017}, so the full ES should be measurable in the same way. Due to the fact that the entire many-body MPS can efficiently be reconstructed from the local reduced density matrices determined by usual quantum tomography \cite{Cramer2010}, such a method dramatically reduces the resource cost from exponential scaling to linear scaling with respect to the system size. Note that the MPS tomography can be applied to arbitrary quantum many-body systems with interactions or/and disorder, as long as the many-body wave function can efficiently be represented by the MPS.

\end{document}